%% file: redstars.tex
\documentclass[apj,onecolumn]{emulateapj}
\usepackage{graphicx}
\usepackage{array}
\usepackage{longtable}

\usepackage{natbib}
\begin{document}

\input{title}

\input{introduction}

\input{census}

\input{population}

\input{discussion}

\input{conclusion}

\bibliographystyle{apj}

\bibliography{bibliography}

\newpage

\input{appendix}

\end{document}

%% file: title.tex
\title{Census of Self-Obscured Massive Stars in Nearby Galaxies with {\em Spitzer}: \\\
Implications for Understanding the Progenitors of SN~2008S-Like Transients} 
\input{authors} 
\begin{abstract}
\input{abstract} 
\end{abstract} 
\keywords{supernovae: general, individual (SN 2008S) 
--- galaxies: individual (M33, NGC~300, M81, NGC~6946) 
--- catalogs}

\maketitle

%% file: authors.tex
\author{Rubab~Khan\altaffilmark{1},
K.~Z.~Stanek\altaffilmark{1,2},
J.~L.~Prieto\altaffilmark{3,4},
C.~S.~Kochanek\altaffilmark{1,2},
Todd~A.~Thompson\altaffilmark{1,2,5}, 
J.~F.~Beacom\altaffilmark{6,1,2},
}

\altaffiltext{1}{Dept.\ of Astronomy, The Ohio State University, 140
W.\ 18th Ave., Columbus, OH 43210; khan, kstanek, ckochanek,
thompson@astronomy.ohio-state.edu}

\altaffiltext{2}{Center for Cosmology and AstroParticle Physics, 
The Ohio State University, 191 W.\ Woodruff Ave., Columbus, OH 43210}

\altaffiltext{3}{Carnegie Observatories, 813 Santa Barbara Street, 
Pasadena, CA 91101; jose@obs.carnegiescience.edu}

\altaffiltext{4}{Hubble, Carnegie-Princeton Fellow.}

\altaffiltext{5}{Alfred P. Sloan Foundation Fellow.}

\altaffiltext{6}{Dept.\ of Physics, The Ohio State University, 191 W.\
Woodruff Ave., Columbus, OH 43210; beacom@mps.ohio-state.edu}

\shorttitle{Self-Obscured Massive Stars}

\shortauthors{Khan et al.~2010}

%% file: abstract.tex
\label{sec:abstract}

A new link in the causal mapping between massive stars and potentially
fatal explosive transients opened with the 2008 discovery of the
dust-obscured progenitors of the luminous outbursts in NGC~6946 and
NGC~300. Here we carry out a systematic mid-IR photometric search for
massive, luminous, self-obscured stars in four nearby galaxies: M33,
NGC~300, M81, and NGC~6946. For detection, we use only the 3.6 $\mu$m
and 4.5 $\mu$m IRAC bands, as these can still be used for multi-epoch
\textit{Spitzer} surveys of nearby galaxies ($\lesssim10$\, Mpc). We
combine familiar PSF and aperture-photometry with an innovative
application of image subtraction to catalog the self-obscured massive stars in these
galaxies. In particular, we verify that stars analogous to the
progenitors of the NGC~6946 (SN~2008S) and NGC~300 transients
are truly rare in all four galaxies: their number may be as low as
$\sim1$ per galaxy at any given moment. This result empirically
supports the idea that the dust-enshrouded phase is a very short-lived
phenomenon in the lives of many massive stars and that these objects
constitute a natural extension of the AGB sequence. We also provide
mid-IR catalogs of sources in NGC~300, M81, and NGC~6946.

%% file: introduction.tex
\section{Introduction}
\label{sec:introduction}

The conventionally-understood fate of massive ($M > 8M_\odot$) stars
is death by a core collapse that produces a neutron star
or black hole, and an outgoing shock wave that successfully ejects the
overlying stellar envelope to create a bright Type~II or Ib/c
supernova (SN). Given the detection of a SN-like optical event, the
next challenge is to characterize the progenitor star in order to
understand the last years of the life of a massive star. A more direct
method for understanding the relation between massive stars and their
transients is to simply catalog all the massive stars in the local
universe ($D\lesssim10$\,Mpc) and then determine their individual
fates. This is particularly important because we lack observational
constraints on the number of failed SN that form a black hole
without a SN-like optical
transient~\citep[e.g.,][]{ref:Kochanek_2008,ref:Smartt_2009}, and
because the boundary between SN and other luminous transients
of massive stars is
uncertain~\citep[e.g.,][]{ref:Chiosi_1986,ref:Woosley_2002}. While
surveys for bright optical transients in the local universe are
well-developed~\citep[e.g.,][]{ref:Li_2001}, a complete census of
massive stars in nearby galaxies are relatively
recent~\citep{ref:Massey_2006} and efforts to study their fates have
just started~\citep{ref:Kochanek_2008}. Despite the technical
challenges required by the depth, area, and cadence of the
observations, these surveys are critical for our understanding of the
correspondences between massive stars and their end states. The
long-term promise of these surveys is to produce a catalog where the
characteristics (luminosity, mass, binarity, winds, etc.) of the
progenitors of future SN are listed.

A new link in this causal mapping between massive stars and their
explosions opened with the discovery of the dust-obscured progenitors
of the luminous outbursts in NGC~6946
\citep[SN~2008S;][]{ref:Arbour_2008,ref:Prieto_2008a} and
NGC~300~\citep[hereafter NGC 300-2008OT;][]{ref:Monard_2008,ref:Bond_2009,ref:Botticella_2009} in the mid-IR with
\textit{Spitzer}. In both cases, the progenitor was not detected in the
optical, but as a $\sim400$K mid-IR source with luminosity of
$\sim5\times10^{4}L_{\odot}$~\citep{ref:Prieto_2008a,ref:Prieto_2008b} 
and an implied mass range of $\sim6-15$ $M_{\odot}$~\citep{ref:Prieto_2008a,ref:Botticella_2009,ref:Berger_2009,ref:Gogarten_2009}.
The dust temperature and the stringent upper limits on the optical
fluxes~\citep{ref:Prieto_2008b,ref:Botticella_2009,ref:Berger_2009} demonstrated that
both objects were enshrouded by a dusty wind. Whether progenitors of
some massive star outburst or SN, the fact that these
progenitors are completely self-obscured by dust implies that any
complete census of the progenitors of luminous transients now requires
mid-IR data as well.

In essence, we need a comprehensive survey for bright mid-IR sources
in all nearby galaxies ($\lesssim10$\,Mpc) with (warm) {\it Spitzer},
analogous to the survey proposed by~\citet{ref:Kochanek_2008} in the
optical. For a first look, \citet{ref:Thompson_2008} searched in M33
for deeply-embedded point sources that had similar properties to the
progenitors of the transients in terms mid-IR colors, luminosities, and
variability in archival {\it Spitzer} observations of M33, and
detected only 18 sources with any similarity, and even fewer (one or two) as extreme in color and
luminosity. They argued that transients such as SN~2008S and
NGC~300-2008OT constitute a class of their own and are relatively common (of
order $\sim10$\% of the SN rate), although the progenitors of
this class are extremely rare among a galaxy's massive stars at a
given time. Thus, \citet{ref:Thompson_2008} argued that a significant fraction of all massive stars must undergo a
dust-enshrouded phase within $\lesssim10^4$\,years prior to some kind
of explosion.

Our understanding of the role of dust and variability in the final
stages of the evolution of massive stars is surprisingly limited,
despite it being a powerful probe of the physics of their last
days. The goal of our project is to expand
the~\citet{ref:Thompson_2008} survey to catalog the population of
obscured massive stars and to identify the sub-population of massive
stars from which these progenitors emerge. We use only the 3.6 $\mu$m
and 4.5 $\mu$m IRAC bands, as they can be used in a future multi-epoch
\textit{Spitzer} survey of all nearby galaxies ($\lesssim10$\,Mpc)
lacking the necessary archival data. We conduct our search in four
galaxies: M33 \citep[D\,$\simeq0.96$\,Mpc,][]{ref:Bonanos_2006},
NGC~300 \citep[D\,$\simeq1.9$\,Mpc,][]{ref:Gieren_2005}, M81
\citep[D\,$\simeq3.6$\,Mpc,][]{ref:Saha_2006}, and NGC~6946
\citep[D\,$\simeq5.6$\,Mpc,][]{ref:Sahu_2006}, so that we can compare
our results with~\citet{ref:Thompson_2008} and test our search method
in relatively distant and crowded galaxies, where a novel application
of image subtraction helps us overcome the poor spatial resolution of
\textit{Spitzer} compared to the stellar crowding. We will also examine 
these populations in the LMC and SMC using the published catalogs of 
\citet{ref:Blum_2006} and \citet{ref:Bolatto_2007}, respectively. These 
galaxies span a broad range in distance, luminosity, star formation rate 
(SFR) and metallicity.

The paper is organized as follows. Section~\ref{sec:census} describes
our methodology for identifying obscured massive stars in nearby
galaxies in detail. Section~\ref{sec:population} discusses the nature
of this stellar population using the color-magnitude diagrams (CMDs) for the
galaxies and spectral energy distributions (SEDs) for the individual obscured
stars. Section~\ref{sec:discussion} considers the implications of the
findings for our understanding of the nature of progenitors of
SN~2008S-like transients and the arguments for conducting a
multi-epoch survey of all nearby galaxies ($\lesssim10$\,Mpc) with
(warm) {\it Spitzer}. Also, we provide lists of the obscured massive
stars that we identified in the four galaxies, and the mid-IR catalogs
of sources in NGC~300, M81, and NGC~6946. The mid-IR catalog of sources in M33
was published in~\citet{ref:Thompson_2008} (also see~\citet{ref:McQuinn_2007}).

\begin{figure}[p]
\begin{center}
{\includegraphics[angle=0,width=85mm]{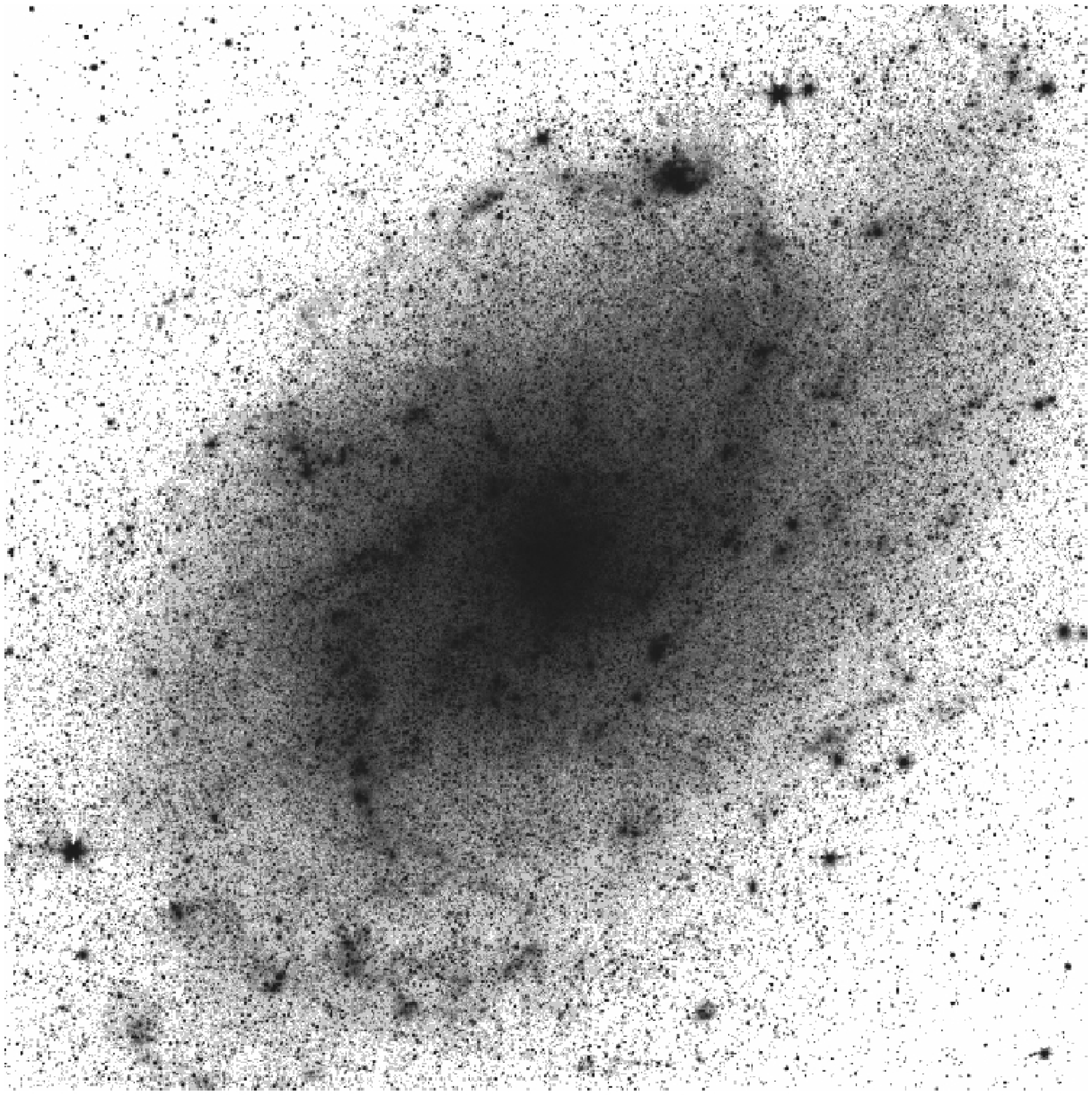}}
{\includegraphics[angle=0,width=85mm]{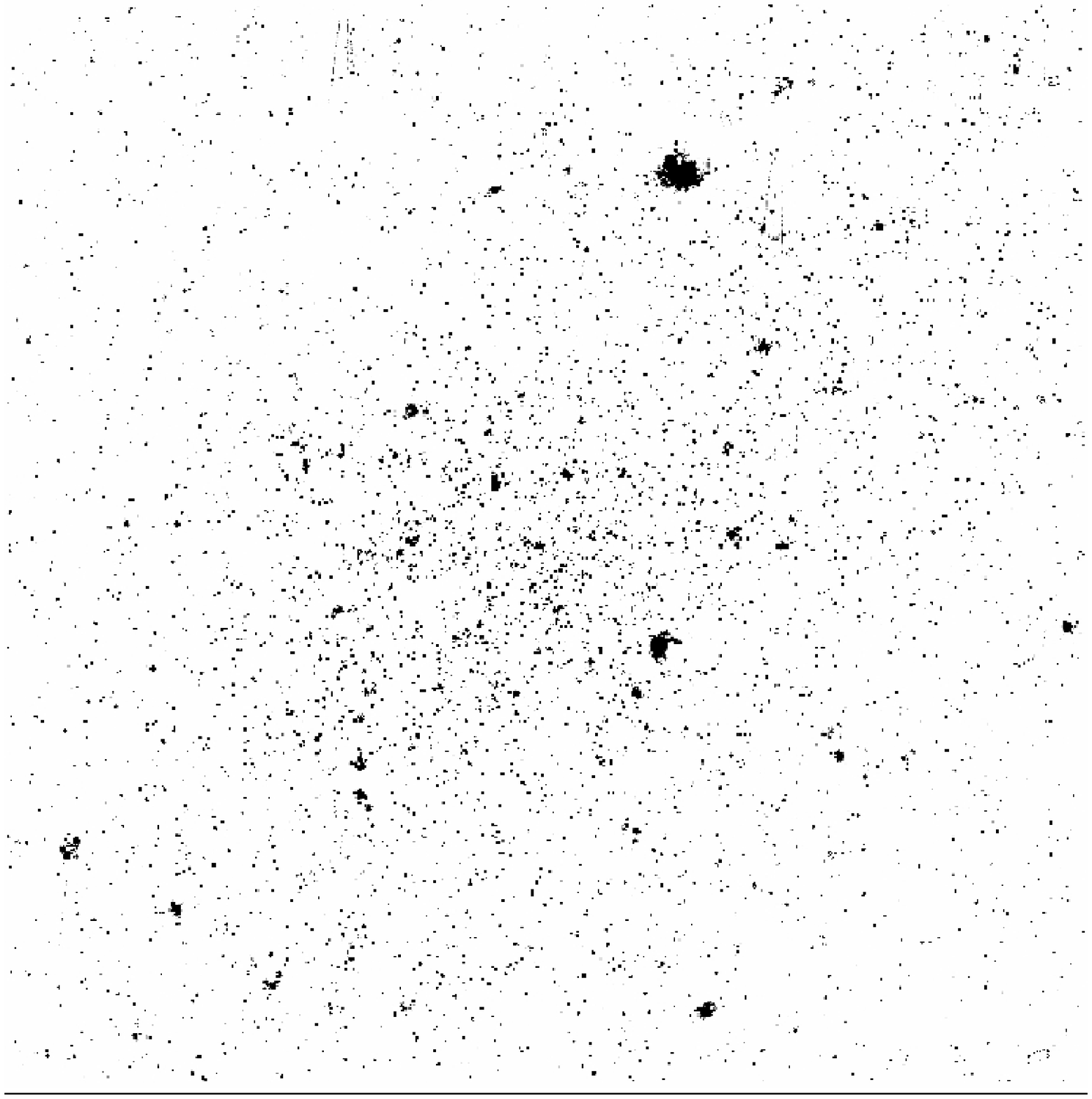}}
\end{center}
\caption{M33 4.5 $\mu$m IRAC image (\textit{left}) and $[3.6]-[4.5]$ image (\textit{right}). The image covers an area of $\approx33'\times33'$ ($1600\times1600$ pixels, with $1\farcs2$/pixel). This difference image is constructed by using image subtraction to scale and subtract the 3.6 $\mu$m image from the 4.5 $\mu$m image including the necessary corrections for the PSF differences. All the normal (non-red) stars ``vanish'' in the [3.6]-[4.5] image leaving the stars with significant dust emission.}
\label{fig:m33}
\begin{center}
{\includegraphics[angle=0,width=85mm]{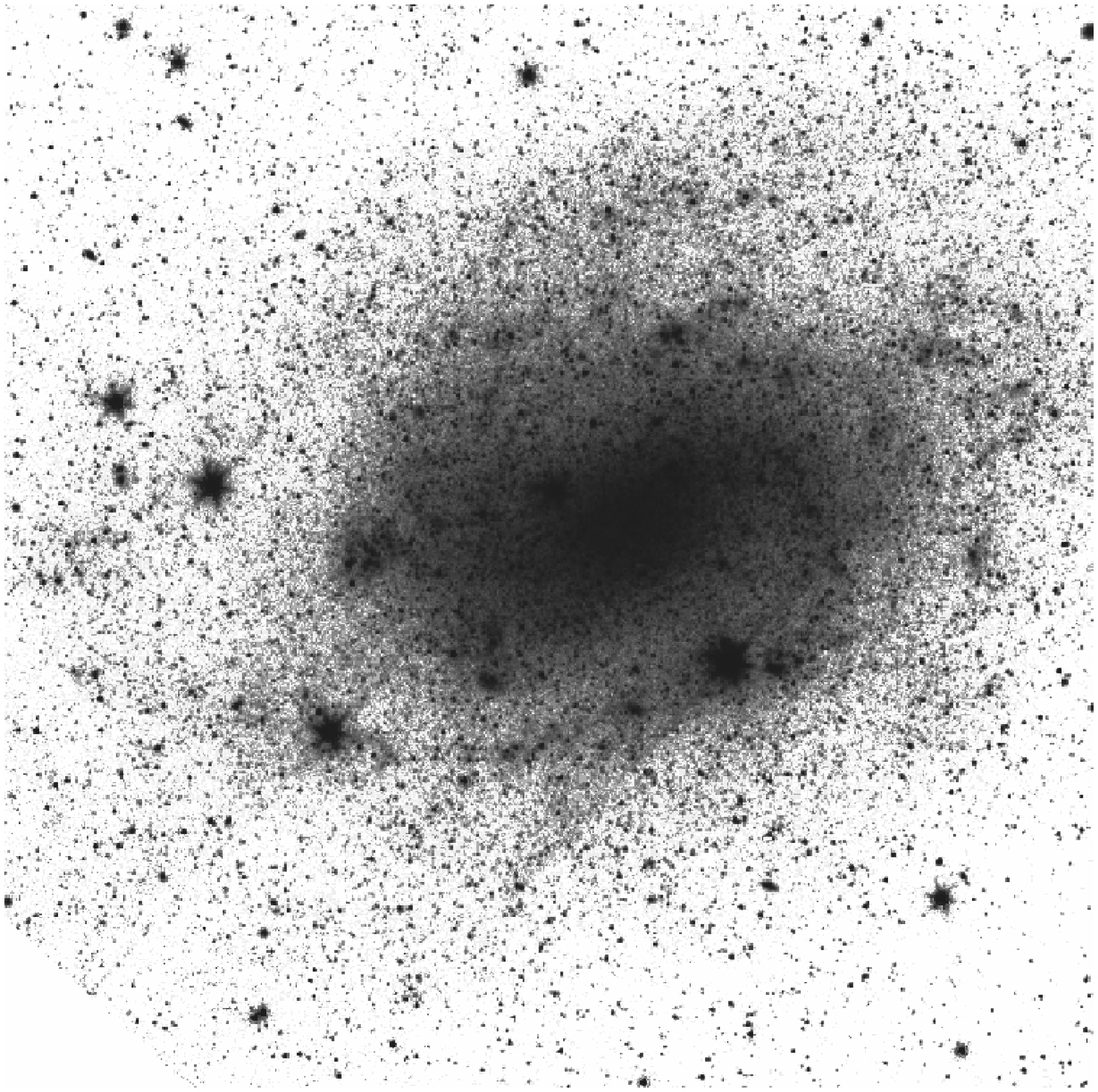}}
{\includegraphics[angle=0,width=85mm]{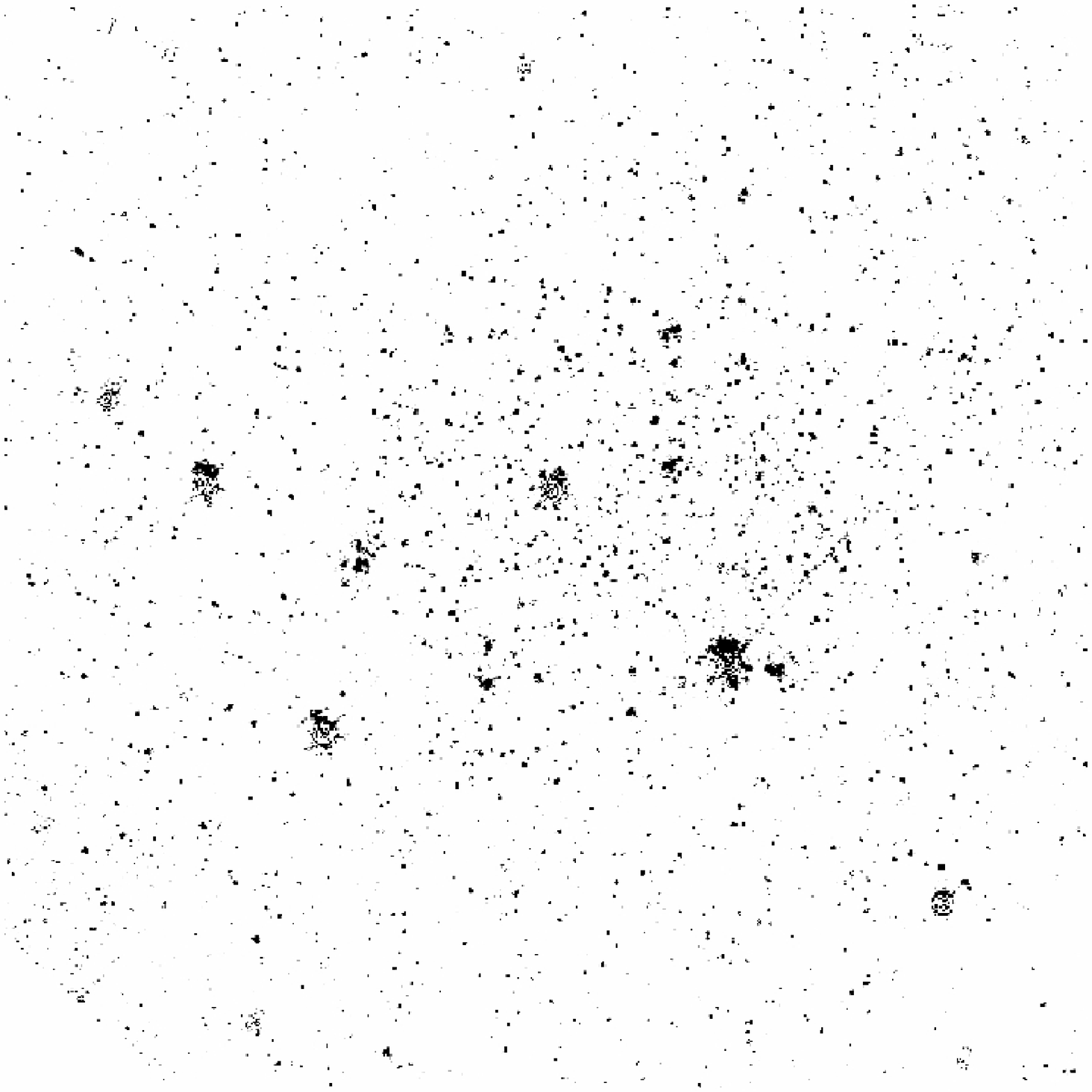}}
\end{center}
\caption{NGC~300 4.5 $\mu$m IRAC image (\textit{left}) and $[3.6]-[4.5]$ image (\textit{right}), as in Figure \ref{fig:m33}. The image covers an area of $\approx15'\times15'$ ($1250\times1250$ pixels with $0\farcs75$/pixel).}
\label{fig:ngc300}
\end{figure}

\begin{figure}[p]
\begin{center}
{\includegraphics[angle=0,width=85mm]{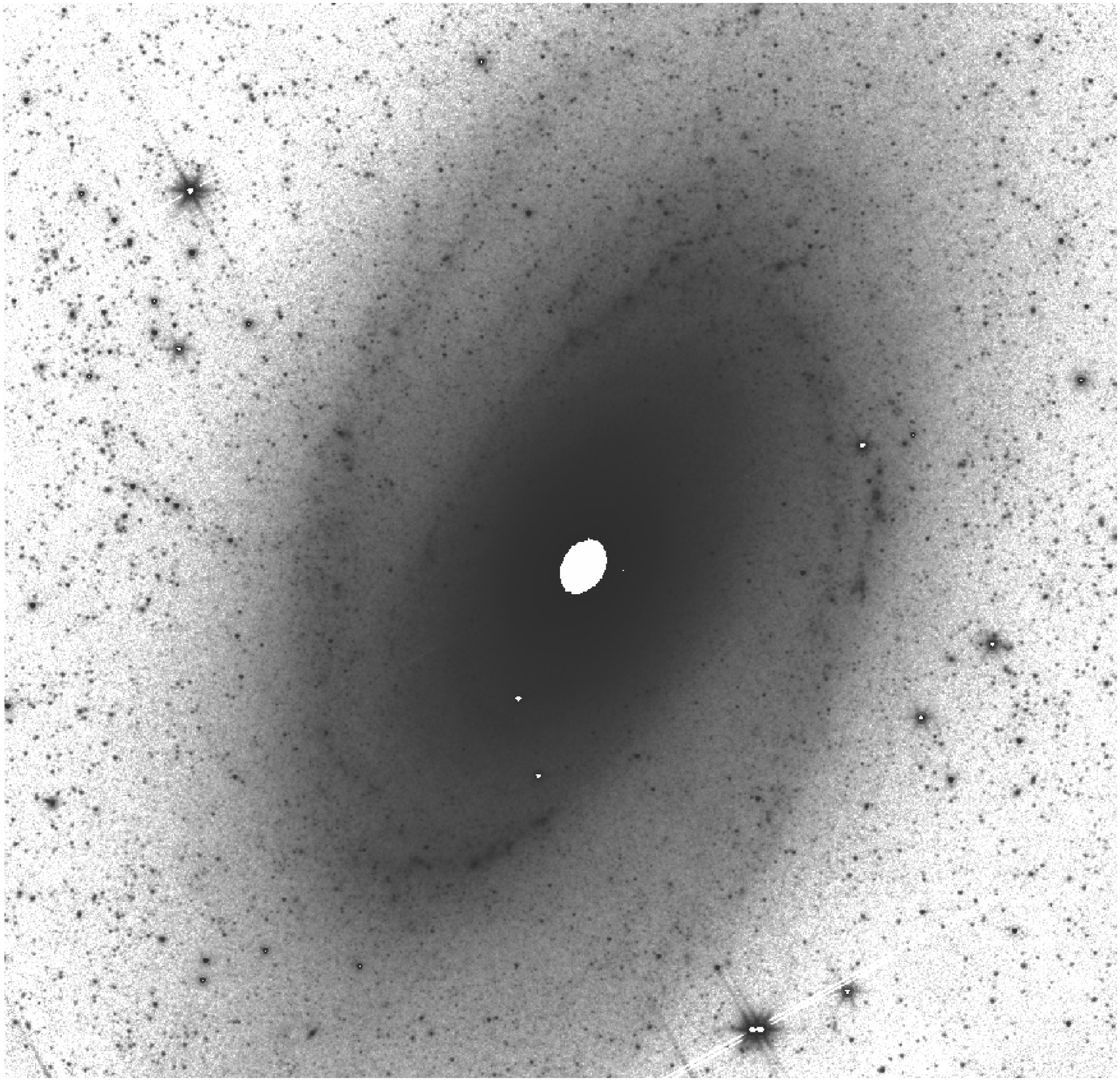}}
{\includegraphics[angle=0,width=85mm]{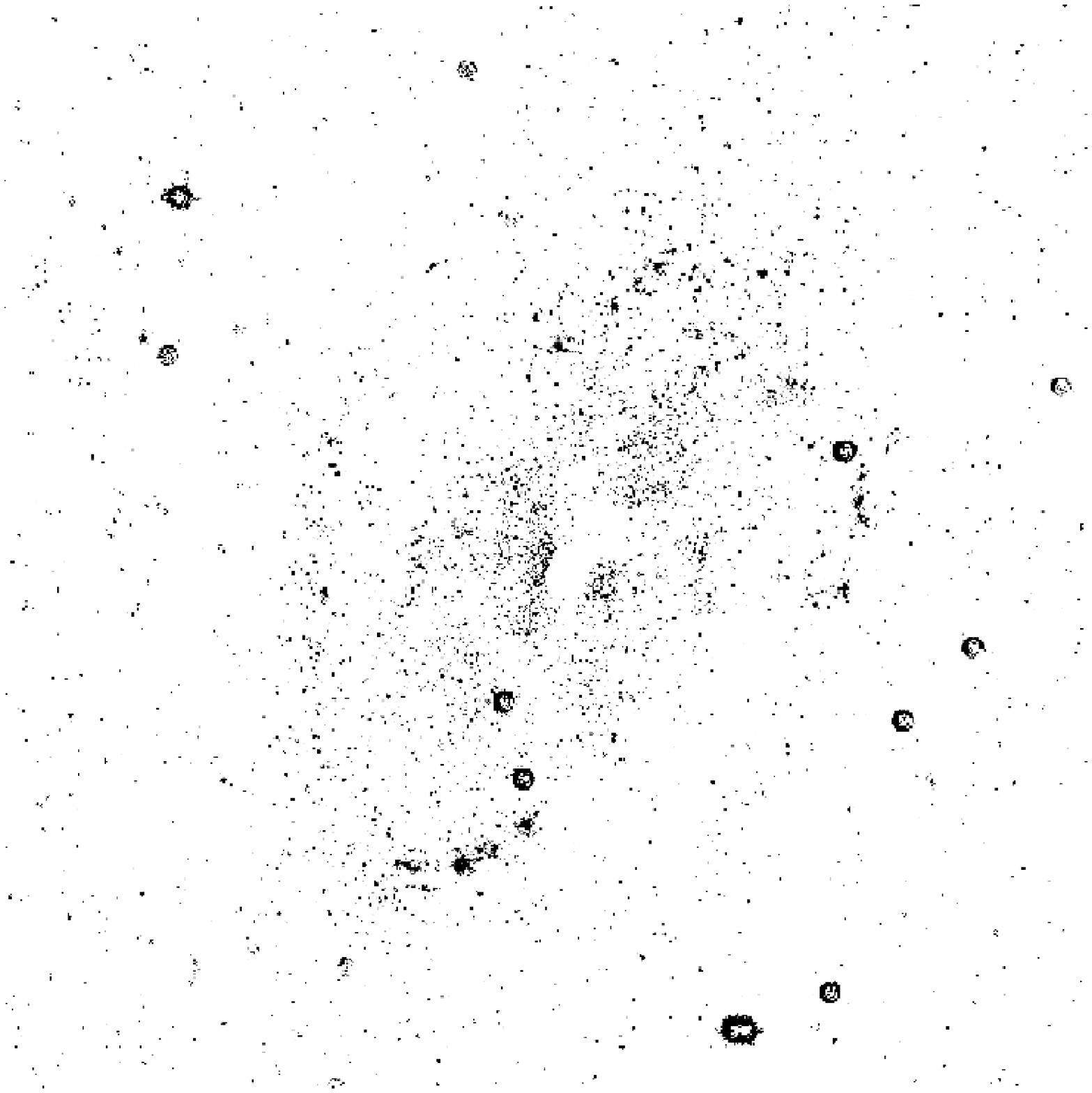}}
\end{center}
\caption{M81 4.5 $\mu$m IRAC image (\textit{left}) and $[3.6]-[4.5]$ image (\textit{right}), as in Figure \ref{fig:m33}. The image covers an area of $\approx18'\times18'$ ($1450\times1450$ pixels with $0\farcs75$/pixel).
The saturated center of M81 has been masked for data reduction purposes. }
\label{fig:m81}
\begin{center}
{\includegraphics[angle=0,width=85mm]{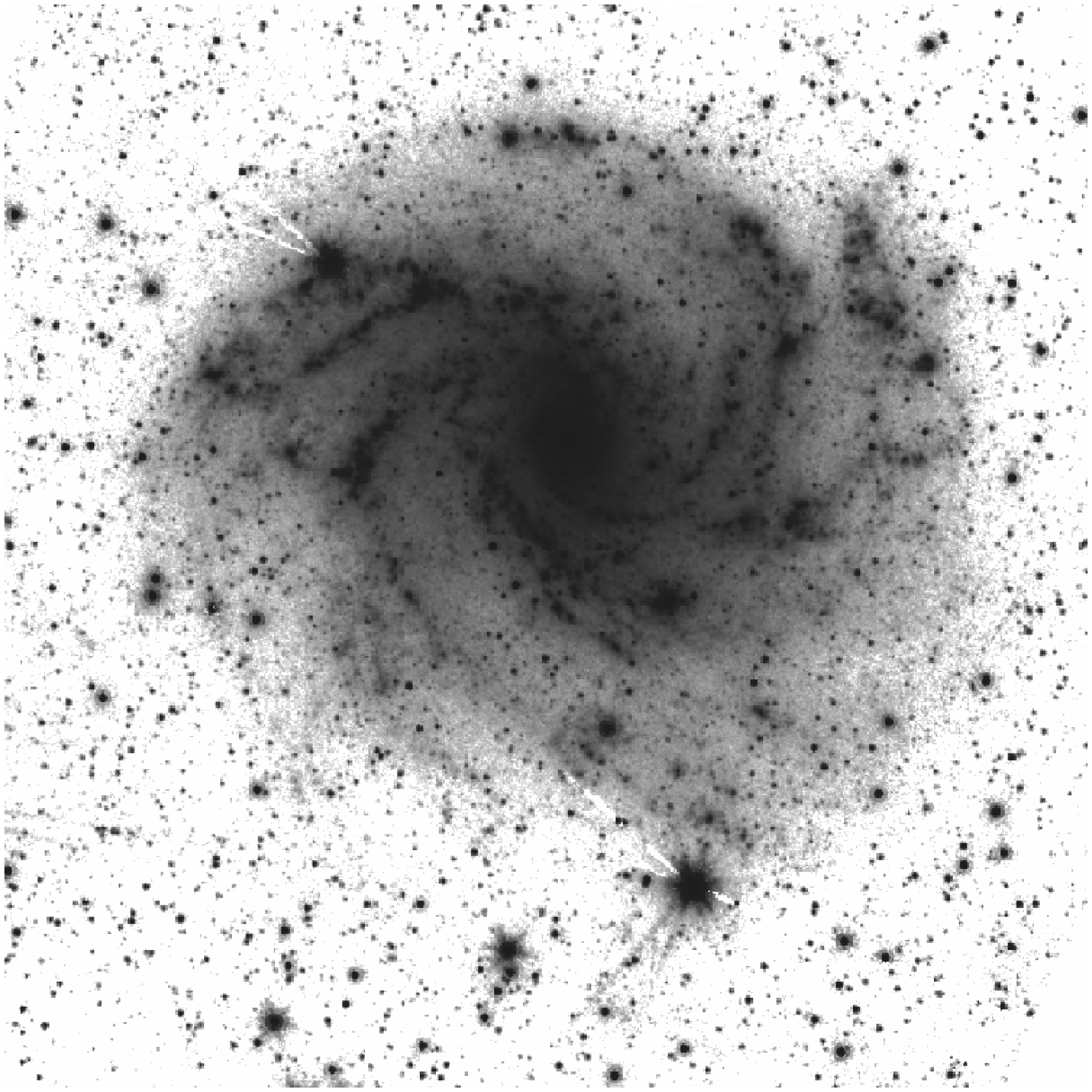}}
{\includegraphics[angle=0,width=85mm]{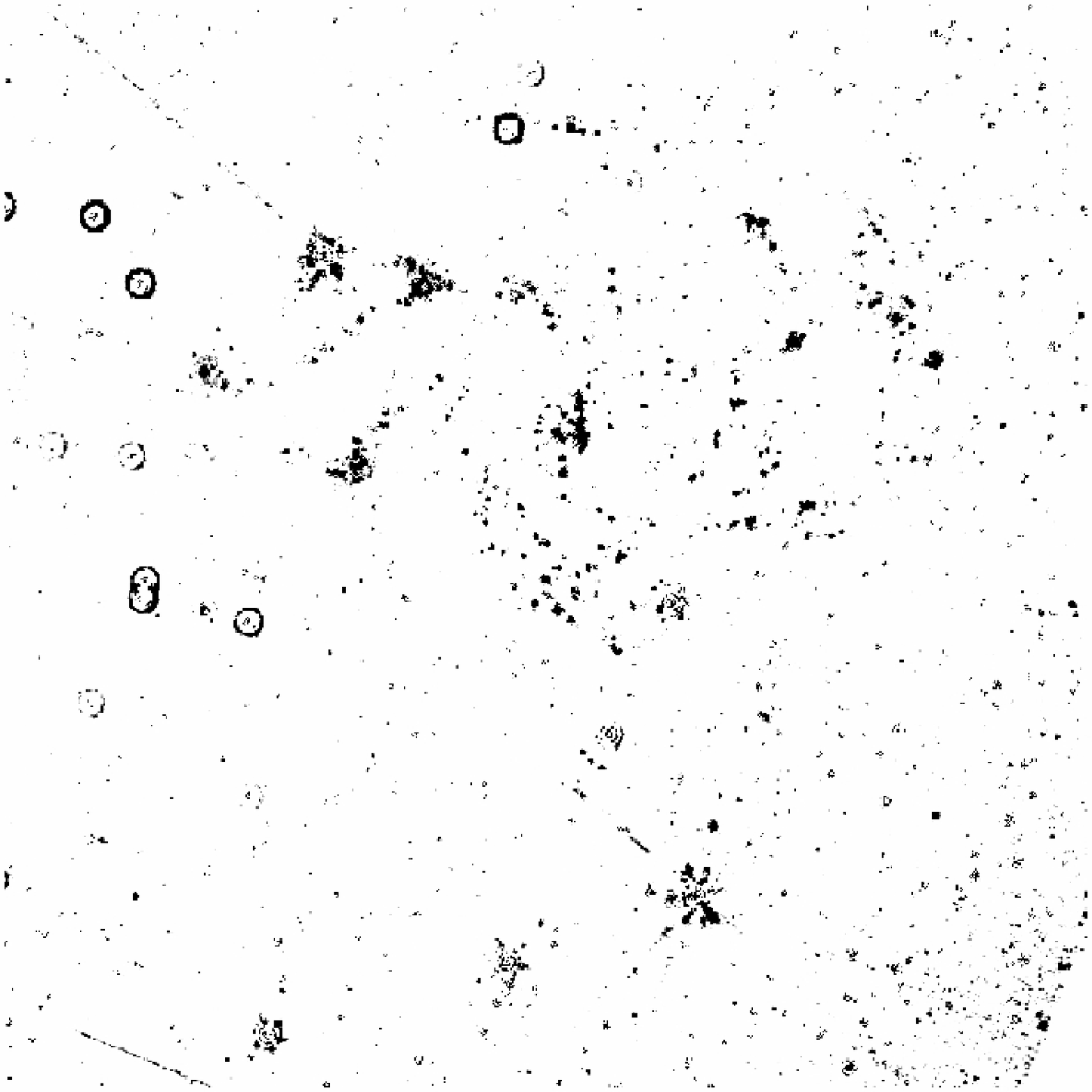}}
\end{center}
\caption{NGC~6946 4.5 $\mu$m IRAC image (\textit{left}) and $[3.6]-[4.5]$ image (\textit{right}), as in Figure \ref{fig:m33}. The image covers an area of $\approx12'\times12'$ ($1000\times1000$ pixels with $0\farcs75$/pixel). The 4.5 $\mu$m  SINGS archival image contains many artifacts that significantly affect the subtracted image. For example, bright stars show bright ``halos'' in the 4.5 $\mu$m image that appear as rings in the subtracted image.}
\label{fig:ngc6946}
\end{figure}

%% file: census.tex
\section{Search for Obscured Massive Stars}
\label{sec:census}

In this section, we describe the methodology of our search for
self-obscured massive stars in archival \textit{Spitzer}
IRAC~\citep{ref:Fazio_2004} data. Our goal is to carry out
this inventory and demonstrate how it can be done despite the crowding
problems created by the limited spatial resolution of \textit{Spitzer}
at greater distances. For M33, we used the six co-added epochs of data
from~\citet{ref:McQuinn_2007} that were used
by~\citet{ref:Thompson_2008}. For NGC~300, we used the data collected by 
the Local Volume Legacy (LVL) Survey~\citep{ref:Dale_2009}. For NGC~6946 
and M81 we used the data collected by the SINGS Legacy Survey~\citep{ref:Kennicutt_2003}. 
Our images of the galaxies are selected regions of the full 
mosaics available for each galaxy. The M33
image covers an area of $\approx33'\times33'$ ($1600\times1600$
pixels, with $1\farcs2$/pixel), the NGC~300 image covers an area of
$\approx15'\times15'$ ($1250\times1250$ pixels with
$0\farcs75$/pixel), the M81 image covers an area of
$\approx18'\times18'$ ($1450\times1450$ pixels with
$0\farcs75$/pixel), and the NGC~6946 image covers an area of
$\approx12'\times12'$ ($1000\times1000$ pixels with
$0\farcs75$/pixel). Figures $1-4$ show the 4.5 $\mu$m
images of the four galaxies.

We searched for obscured stars in three stages and order the
identified objects by the stage at which we can fully measure the
source properties. We use the same criteria of $[3.6]-[4.5]>1.5$
magnitude and $M_{4.5}<-10$ magnitude used
by~\citet{ref:Thompson_2008} to select extremely red and bright
objects (Extreme Asymptotic Giant Branch or EAGB objects) similar to
the SN~2008S and NGC~300-2008OT progenitors. First, we
searched for objects directly identifiable in both the 3.6 $\mu$m and
4.5 $\mu$m images. Second, we searched for objects detectable at 4.5
$\mu$m but not at 3.6 $\mu$m. Finally, in a novel application of
difference imaging methods, we searched for obscured stars not
directly detectable in either the 3.6 $\mu$m or 4.5 $\mu$m IRAC
bands. For nearer galaxies, we expect most red sources to be
detectable in either or both of the IRAC bands. But for the more
distant galaxies, the difference imaging method helps us identify
confused sources in crowded images.

All normal stars have the same mid-IR color (Vega
$[3.6]-[4.5]\simeq0$), because of the Rayleigh-Jeans
tails of their spectra. Stars with dusty envelopes lie off the main
stellar locus at color 0 towards redder colors. This means that if we
use difference imaging methods to match the 3.6 $\mu$m image to the
flux scale and point spread function (PSF) structure of the 4.5 $\mu$m image and then
subtract, all the normal stars ``vanish'' to leave us only with the
stars having significant dust emission. In practice, we use the ISIS
image subtraction software package~\citep{ref:Alard_1998} following
the procedures from~\cite{ref:Hartman_2004}. We use the 4.5 $\mu$m image 
as the reference from which the 3.6 $\mu$m image is subtracted. 
ISIS automatically sets the flux scaling and backgrounds 
to optimally subtract the two bands, and utilizes a spatially variable 
kernel to model the PSF differences. Figures $1-4$ show the
results of this procedure for our target galaxies. The result is
visibly stunning: almost all the stars ``vanish'' to leave us with a
clean image of the red self-obscured stars. Now we can identify and
accurately measure positions for red sources that could otherwise be
confused due to crowding.

\begin{figure}[p]
\begin{center}
{\includegraphics[angle=0,width=160mm]{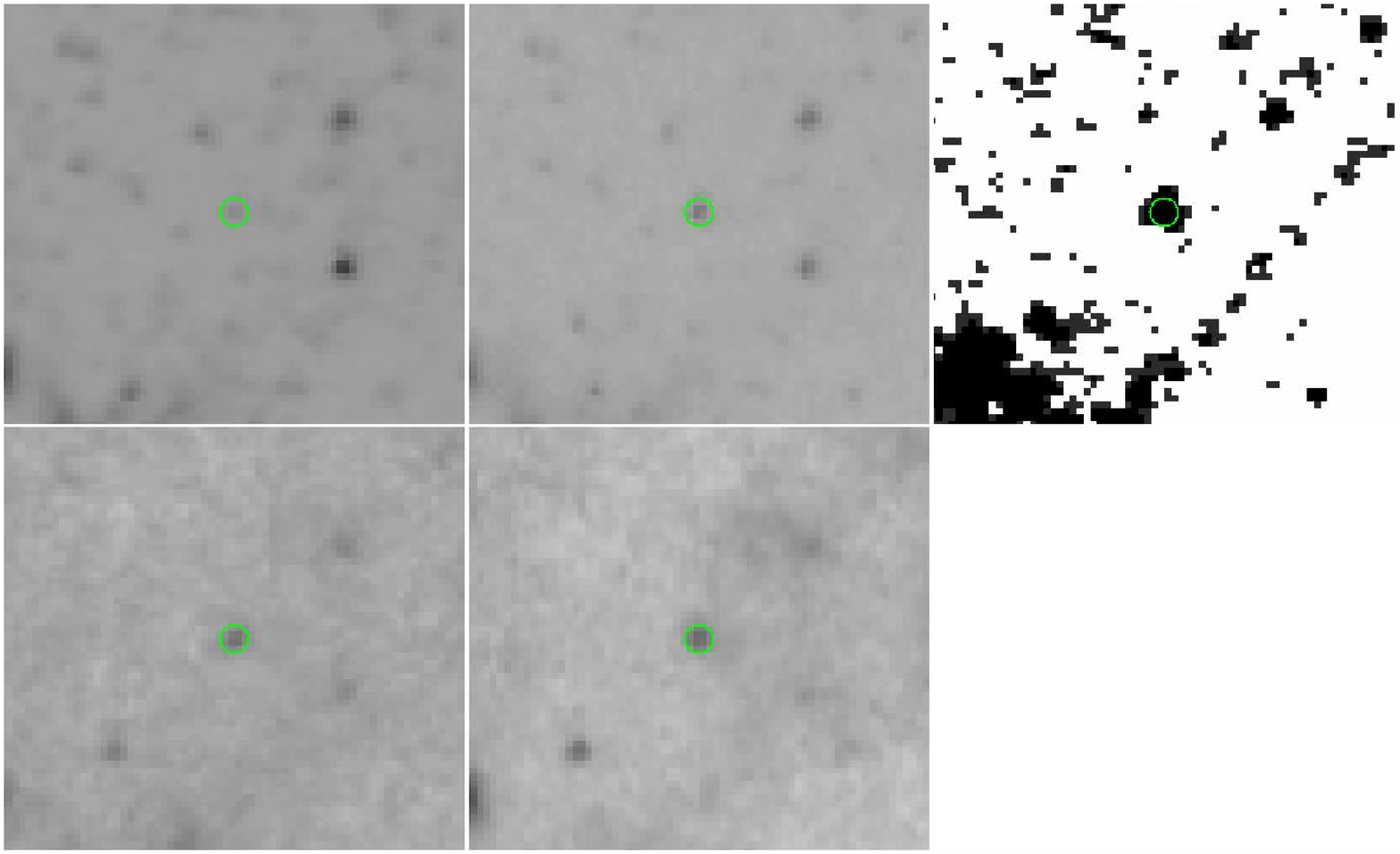}}
\end{center}
\caption{A Class--A object in NGC~300 visible at both 3.6 $\mu$m (\textit{top left}) and 4.5 $\mu$m (\textit{top center}). The $[3.6]-[4.5]$ (\textit{top right}), 5.8 $\mu$m (\textit{bottom left}), and 8.0 $\mu$m (\textit{bottom center}) images are also shown. Each panel is $\sim52\farcs5$ on its sides.}
\label{fig:ngc300_step1}
\begin{center}
{\includegraphics[angle=0,width=160mm]{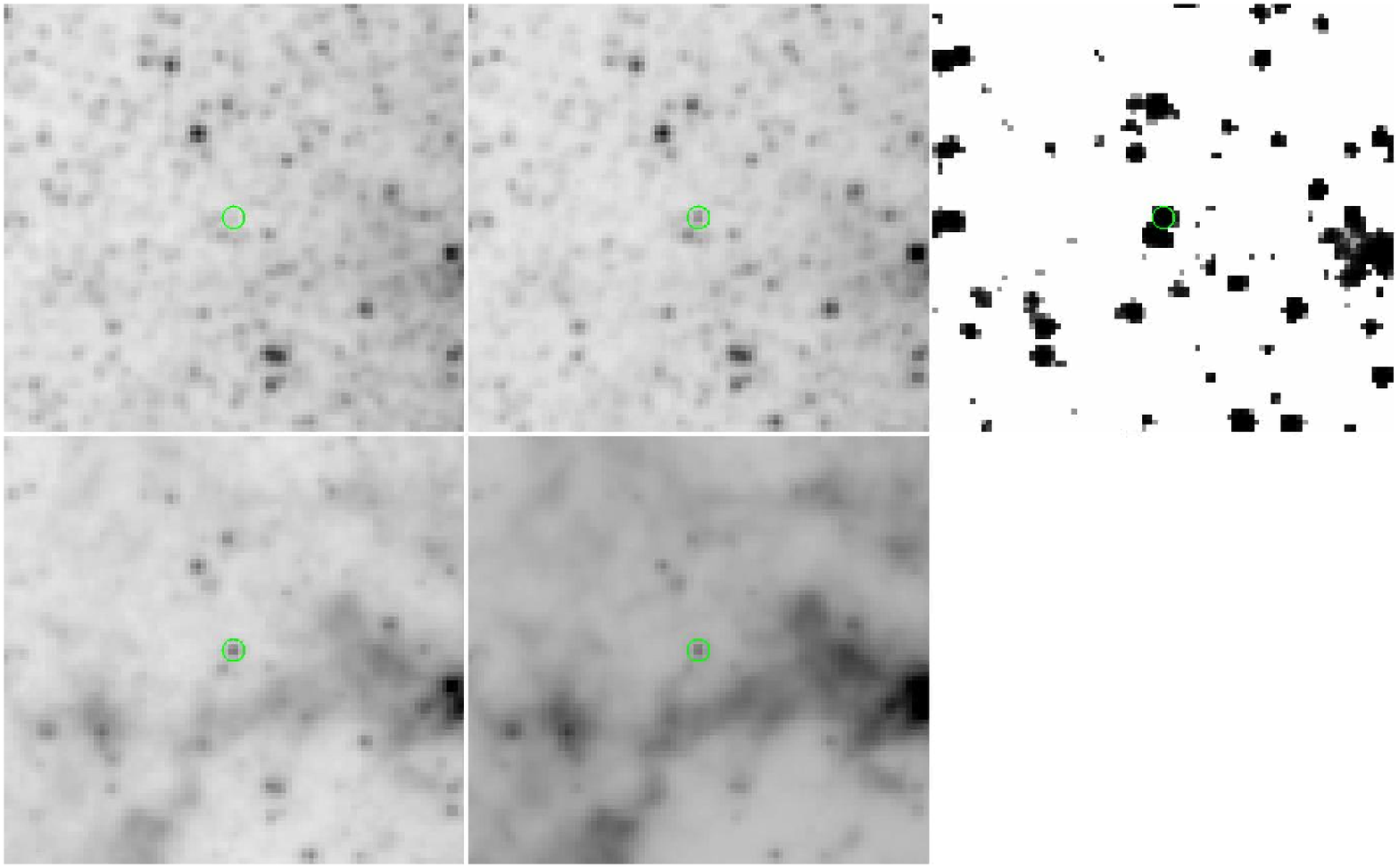}}
\end{center}
\caption{A Class--B object in M33 visible only at 4.5 $\mu$m (\textit{top center}) and not at 3.6 $\mu$m (\textit{top left}). The 3.6 $\mu$m magnitude is determined at the position of the 4.5 $\mu$m source. The $[3.6]-[4.5]$ (\textit{top right}), 5.8 $\mu$m (\textit{bottom left}), and 8.0 $\mu$m (\textit{bottom center}) images are also shown. Each panel is $\sim106\farcs4$ on its sides.}
\label{fig:m33_step2} 
\end{figure}

We define objects directly identifiable in both the 3.6 $\mu$m and 4.5
$\mu$m bands as Class--A objects. This stage should be most successful
for the nearest galaxies, and for sources with significant 3.6 $\mu$m
flux. The procedure followed at this stage is similar to that followed
by~\citet{ref:Thompson_2008}, except that we verify the color limits
for the candidates. Also, for all objects that are not red
enough based on DAOPHOT PSF-fitting methods, we re-estimate the
magnitudes and colors through aperture-photometry to ensure that we do not miss any candidate
objects due to small measurement variations. Figure~\ref{fig:ngc300_step1} shows a Class--A object
detected in NGC~300.

We define objects detectable at 4.5 $\mu$m but not at 3.6 $\mu$m as Class--B objects. For point sources that had a $3\sigma$ detection in the 4.5 $\mu$m image based on PSF-photometry but lacked a 3.6 $\mu$m counterpart within a 0.5 pixel matching radius, we used aperture-photometry at the position of the 4.5 $\mu$m source to estimate or set limits on the 3.6 $\mu$m magnitude and color ($[3.6]-[4.5]$). Figure~\ref{fig:m33_step2} shows a Class--B object detected in M33. 

\begin{figure}[t]
\begin{center}
{\includegraphics[angle=0,width=160mm]{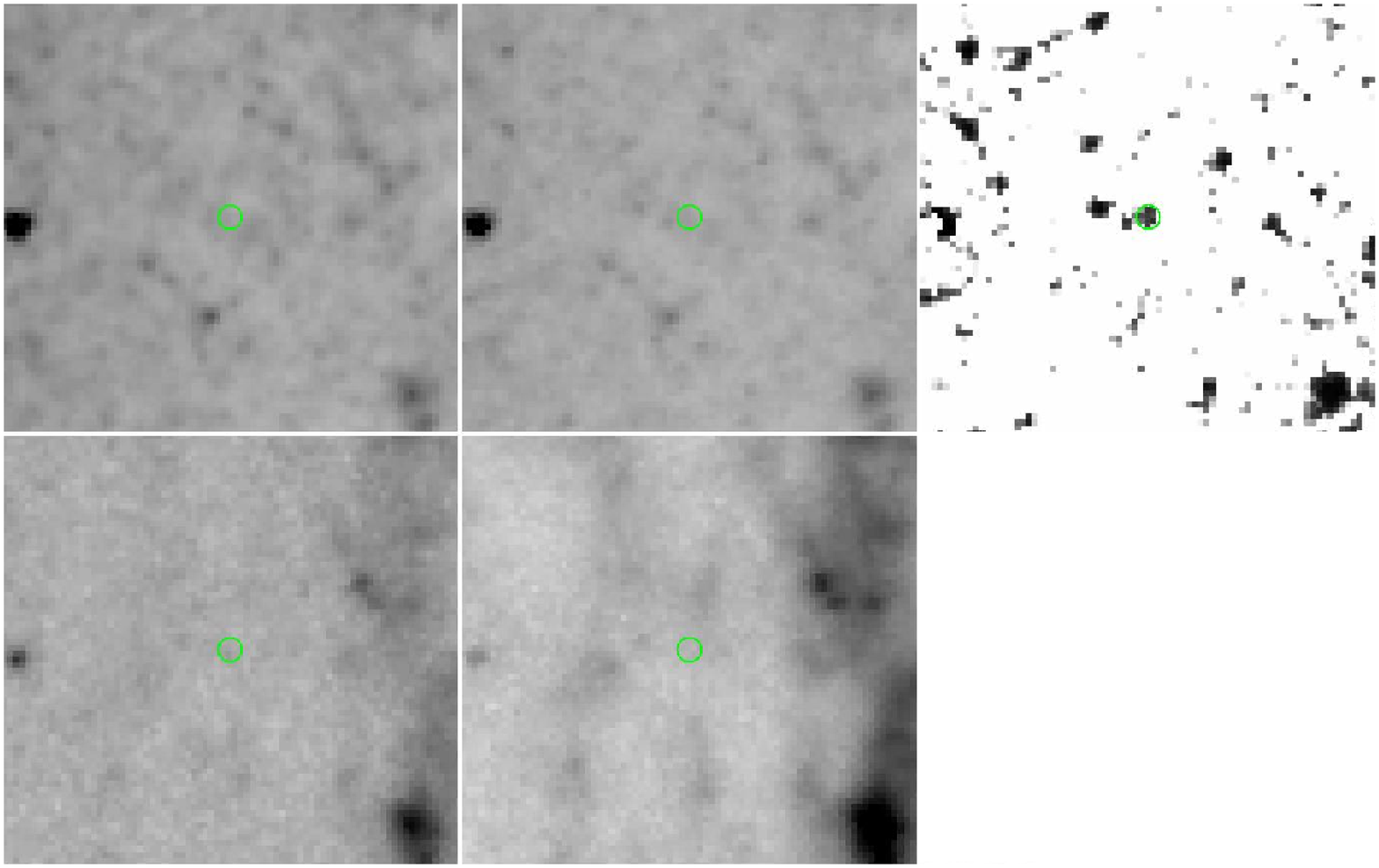}}
\end{center}
\caption{A Class--C object in M81 found only in the $[3.6]-[4.5]$ image (\textit{top right}) but at neither 3.6 $\mu$m (\textit{top left}) nor 4.5 $\mu$m (\textit{top center}). The magnitudes are determined through aperture-photometry at the location identified in the differenced image. The 5.8 $\mu$m (\textit{bottom left}) and 8.0 $\mu$m (\textit{bottom center}) images are also shown. Each panel is $\sim60\farcs0$ on its sides.}
\label{fig:m81_step3}
\end{figure}

We define objects detectable only in the wavelength differenced image (hereafter the ``$[3.6]-[4.5]$ image'') as Class--C objects. We used DAOPHOT PSF-fitting to identify sources on the $[3.6]-[4.5]$ image. For the detected objects lacking a counterpart in the 4.5 $\mu$m catalog within a 1 pixel matching radius, we use aperture-photometry on both the 3.6 $\mu$m and 4.5 $\mu$m images at the source position in the differenced image to estimate magnitudes and colors. Figure~\ref{fig:m81_step3} shows a Class--C object detected in M81. For comparison, we show the SN~2008S and NGC~300-2008OT in Figures \ref{fig:n300_trans} and \ref{fig:sn2008s}. These are detected as Class--B and Class--A objects, respectively, in our blind searches of NGC~6946 and NGC~300, as described in Section \ref{sec:population}.

Given this broad outline, we now describe the specific technical details of how we performed the photometric measurements at the various stages of the search and then verified the properties of the candidates.

We used the DAOPHOT/ALLSTAR PSF-fitting and photometry package~\citep{ref:Stetson_1992} to identify point sources in both the 3.6 $\mu$m and 4.5 $\mu$m IRAC images. The point source catalogs (Tables 5, 7, and 9) are comprised of those sources detected at $>3\sigma$ at 4.5 $\mu$m and with a 3.6 $\mu$m counterpart within 0.5 pixels. We measure the properties of the sources using either DAOPHOT/ALLSTAR to do PSF-photometry or  APPHOT/PHOT to do simple aperture-photometry. The PSF-magnitudes obtained with ALLSTAR were transformed to Vega-calibrated magnitudes using aperture corrections derived from bright stars. For aperture-photometry magnitudes, the aperture correction is determined from the \textit{Spitzer} point response function reference image obtained from the IRAC website\footnote{http://ssc.spitzer.caltech.edu/irac/psf.html}, and we empirically verified that for bright stars the magnitudes determined through PSF and aperture-photometry agree within $\pm0.05$ magnitude. When performing aperture-photometry,  we estimate the local background for each object using APPHOT/PHOT employing a $2\sigma$ outlier rejection procedure in order to exclude sources located in the local sky annulus, and correct for the excluded pixels assuming a Gaussian background distribution.

For every candidate object, we implement a strict detection criteria. We require a $>3\sigma$ detection at 4.5 $\mu$m regardless of the search stage at which an object is detected and how its properties are determined (PSF or aperture-photometry). Initially, we determine the $2\sigma$ limit on $m_{3.6}$ using the local background estimated using APPHOT/PHOT. If the estimate of $m_{3.6}$ is brighter than the $2\sigma$ local background limit, then we treat the $m_{3.6}$ estimate as the measured flux, otherwise we use the $2\sigma$ background estimate as an upper limit on the flux. Thus, we get a $2\sigma$ lower limit on the $[3.6]-[4.5]$ color of the objects for which we do not have reliable measurements at 3.6 $\mu$m. This color limit verification is done identically in every stage of the search. Stars for which only limits could be determined are included in the tables and CMDs, but we do not show their SEDs. We visually inspected all candidate red stars in the 3.6 $\mu$m, 4.5 $\mu$m, and wavelength differenced images and rejected candidates that do not appear likely to be a star. The differenced images are very useful because they cleanly remove most of the crowding, as is apparent in Figures $5-9$.

Due to how we organized the search, we first identify candidates that can be detected it at least one of the 3.6 $\mu$m and 4.5 $\mu$m images before searching for additional objects in the differenced image. Most of the Class--A and the Class--B objects are detected independently in the $[3.6]-[4.5]$ image as well, and the Class--C objects represent only the additional sources that can be hidden by crowding, especially in the more distant galaxies. The progenitor of SN~2008S is an excellent example: although we identified it in NGC~6946 as a Class--B object, Figure \ref{fig:n300_trans} makes it clear that had we missed it at 4.5 $\mu$m, it would be detected without any confusion as a Class--C object. There was a tendency for DAOPHOT to split objects detected in the $[3.6]-[4.5]$ images, presumably due to the non-standard statistical properties of these images.  Generally there was a primary, true detection and a secondary detection offset by roughly a pixel. These off-center detections represented the major source of false-positives for the Class C objects, and we systematically rejected these duplicate detections.

%% file: population.tex
\section{Inventory of Obscured Massive Stars}
\label{sec:population}

Figures $10-13$ present the $m_{4.5}$ versus $[3.6]-[4.5]$
CMDs for each galaxy. For comparison, we
include the progenitors of NGC~300-2008OT and
SN~2008S~\citep{ref:Prieto_2008a,ref:Prieto_2008b} as well as the
$[3.6]-[4.5]>1.5$ and $M_{4.5}<-10$ selection region for extremely red
and bright objects. The candidates that meet this criteria are shown
with symbols indicating the stage of the search at which they were
identified. Where applicable, color limits are indicated by
arrows. Figures $10-13$ also present the mid-IR SEDs of the candidate
objects as compared to the SEDs of the SN 2008S and NGC~300 transient
progenitors. Here we include the 5.8 $\mu$m and 8.0 $\mu$m fluxes as
determined using aperture-photometry for the position of the 4.5
$\mu$m source. Due to significant PAH emissions in these two bands, we
view the aperture-photometry measurements in these bands as less
reliable. The SEDs of some fainter sources show a sharp decline at 5.8
$\mu$m before rising again at 8.0 $\mu$m due to PAH dominated
background contamination. Figure~\ref{fig:all_cmd} shows a summary of
the CMDs of these four galaxies as well as those for the LMC and SMC
from~\citet{ref:Blum_2006} and~\citet{ref:Bolatto_2007},
respectively. Figure~\ref{fig:all_sky} shows the spatial distribution
of the sources. All the objects identified at both 3.6 $\mu$m and 4.5 $\mu$m with
DAOPHOT are reported in Tables \ref{table:ngc300_cat},
\ref{table:m81_cat}, and \ref{table:ngc6946_cat} for NGC~300, M81, and
NGC~6946 respectively. Since~\citet{ref:Thompson_2008} published a
source catalog for M33, we do not publish a new catalog for this
galaxy. In this section, we discuss the results of our search for
obscured massive stars in each of the targeted galaxies. Lists of the
candidates are presented in Tables \ref{table:m33_red},
\ref{table:ngc300_red}, \ref{table:m81_red}, and
\ref{table:ngc6946_red}, sorted by the stage at which we measured the
source properties (Class--A objects first, followed by Class--B objects
and then Class--C objects), for all four galaxies.

\begin{figure}[p]
\begin{center}
{\includegraphics[angle=0,width=160mm]{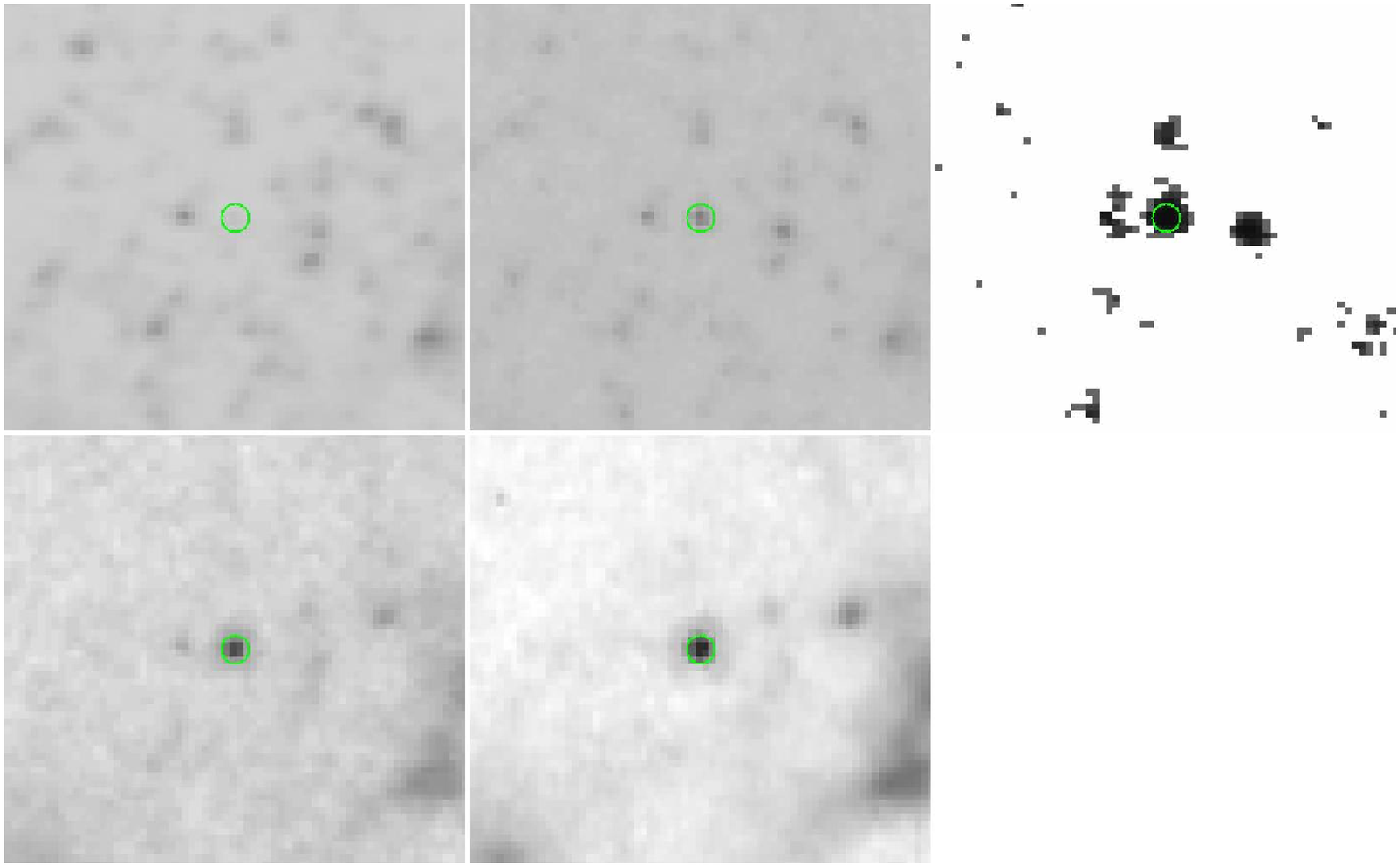}}
\end{center}
\caption{The NGC~300-2008OT progenitor is identified as a Class--A object detected at both 3.6 $\mu$m (\textit{top left}) and 4.5 $\mu$m (\textit{top center}). The $[3.6]-[4.5]$ (\textit{top right}), 5.8 $\mu$m (\textit{bottom left}), and 8.0 $\mu$m (\textit{bottom center}) images are also shown. Each panel is $\sim60\farcs0$ on its sides.}
\label{fig:n300_trans}
\begin{center}
{\includegraphics[angle=0,width=160mm]{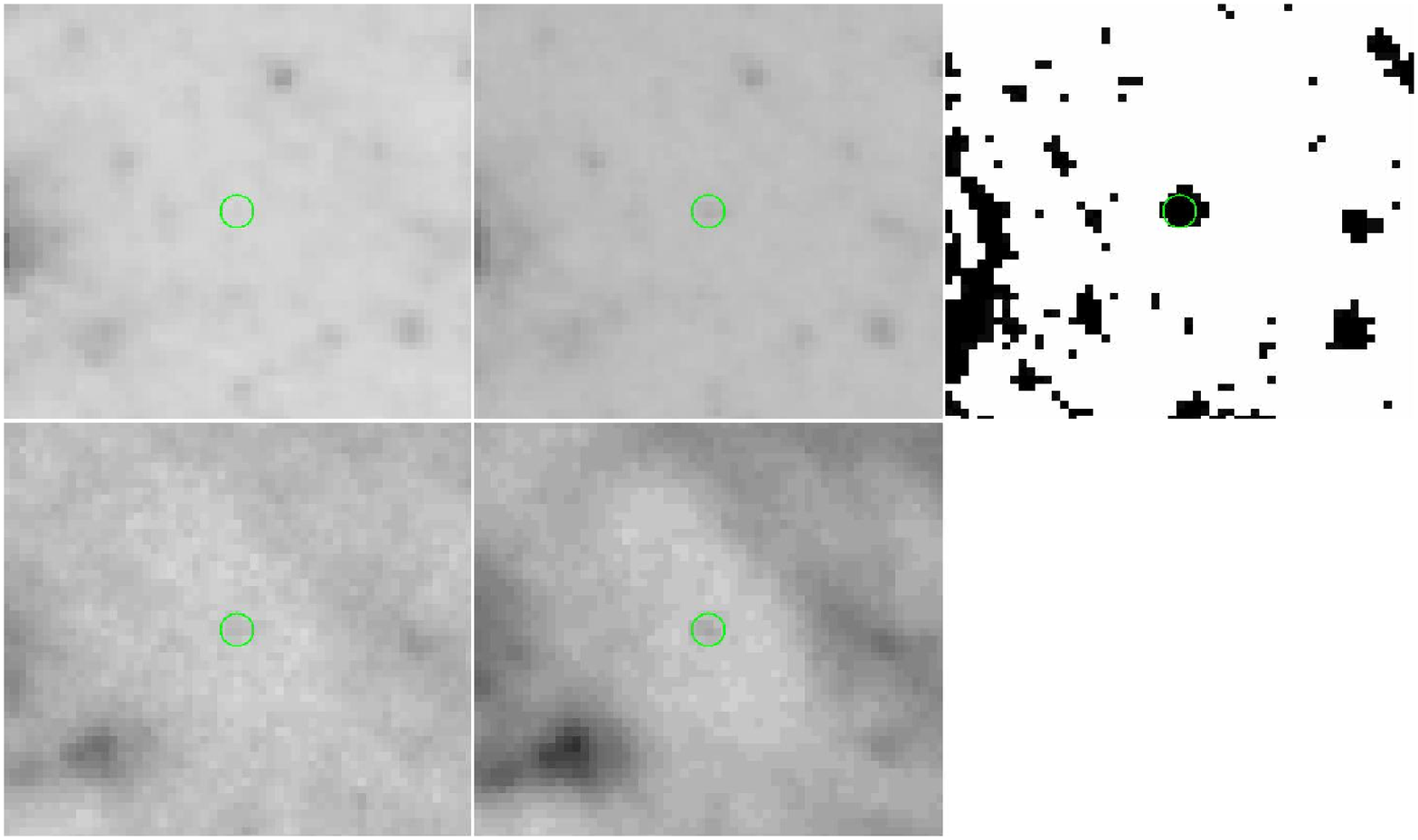}}
\end{center}
\caption{The SN~2008S progenitor, identified in NGC~6946 as a Class--B object, is detected only at 4.5 $\mu$m (\textit{top center})  but not at 3.6 $\mu$m (\textit{top left}). The [3.6]-[4.5] image (\textit{top right}) makes it clear that had we missed it at 4.5 $\mu$m, it would be detected without any confusion as a Class--C object. The 5.8 $\mu$m (\textit{bottom left}), and 8.0 $\mu$m (\textit{bottom center}) images are also shown. Each panel is $\sim41\farcs3$ on its sides.}
\label{fig:sn2008s}
\end{figure}

\subsection{M33}

We first discuss our analysis of M33 \citep[D\,$\simeq0.96$\,Mpc,][]{ref:Bonanos_2006}, the nearest of the targeted galaxies, using
our improved methods. Initially, we identify 15 Class--A objects based
on PSF-photometry exactly following the procedures
of~\citet{ref:Thompson_2008} except where we estimate the color
limits. We identify 4 additional Class--A objects through
aperture-photometry, as well as 3 Class--B objects and 1 Class--C
object. On visual inspection, we reject 1 object in each class,
leaving 20 candidates. These objects include 12 of the 18 stars
identified by~\citet{ref:Thompson_2008} as self-obscured EAGBs. For
one of these objects (Class--B), we could only determine a $2\sigma$
color limit. The only Class--C object was rejected as a duplicate
detection of one of the Class--A objects. Table~\ref{table:m33_red}
presents the candidate EAGB sources. See Table 1 in~\citet{ref:Thompson_2008}
for the mid-IR point source catalog.

Of the 6 stars in~\citet{ref:Thompson_2008} that were not identified
in our present census as candidates, 5 are actually found in
our search (3 in the first stage, 1 in the second stage, 1 in the
third stage), but they are all located close to the color and
magnitude selection boundaries, and small shifts in the color and
magnitude estimates excluded them from the sample. The only object
among the 18 that we missed entirely lay outside our search region,
which is slightly different from that of~\citet{ref:Thompson_2008}. Only 1 source, the same in both searches, is brighter than
the NGC~300 transient progenitor and redder than the SN~2008S
progenitor color limit.

\begin{figure}[p]
\begin{center}
{\includegraphics[angle=0,width=85mm]{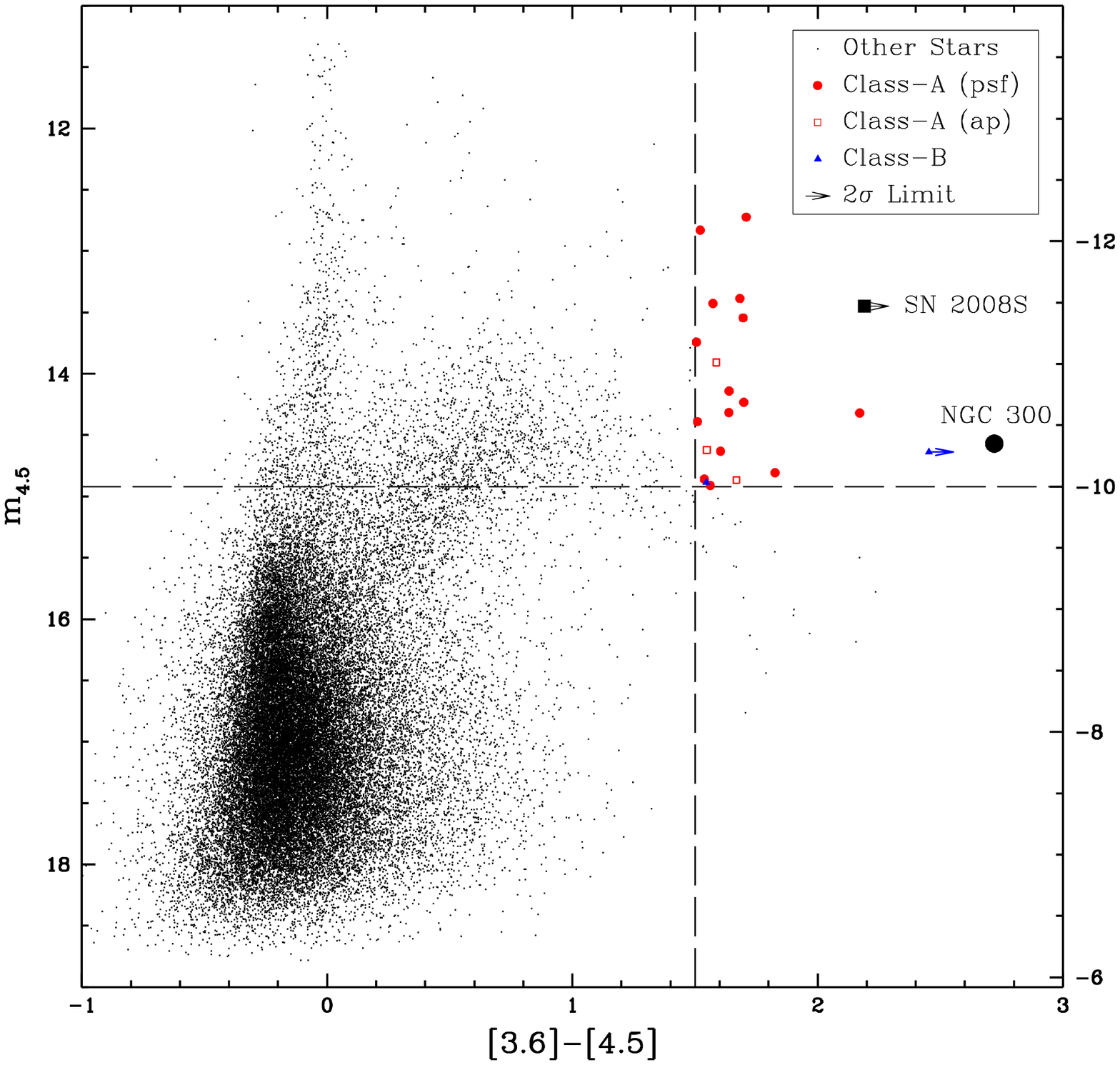}}
{\includegraphics[angle=0,width=85mm]{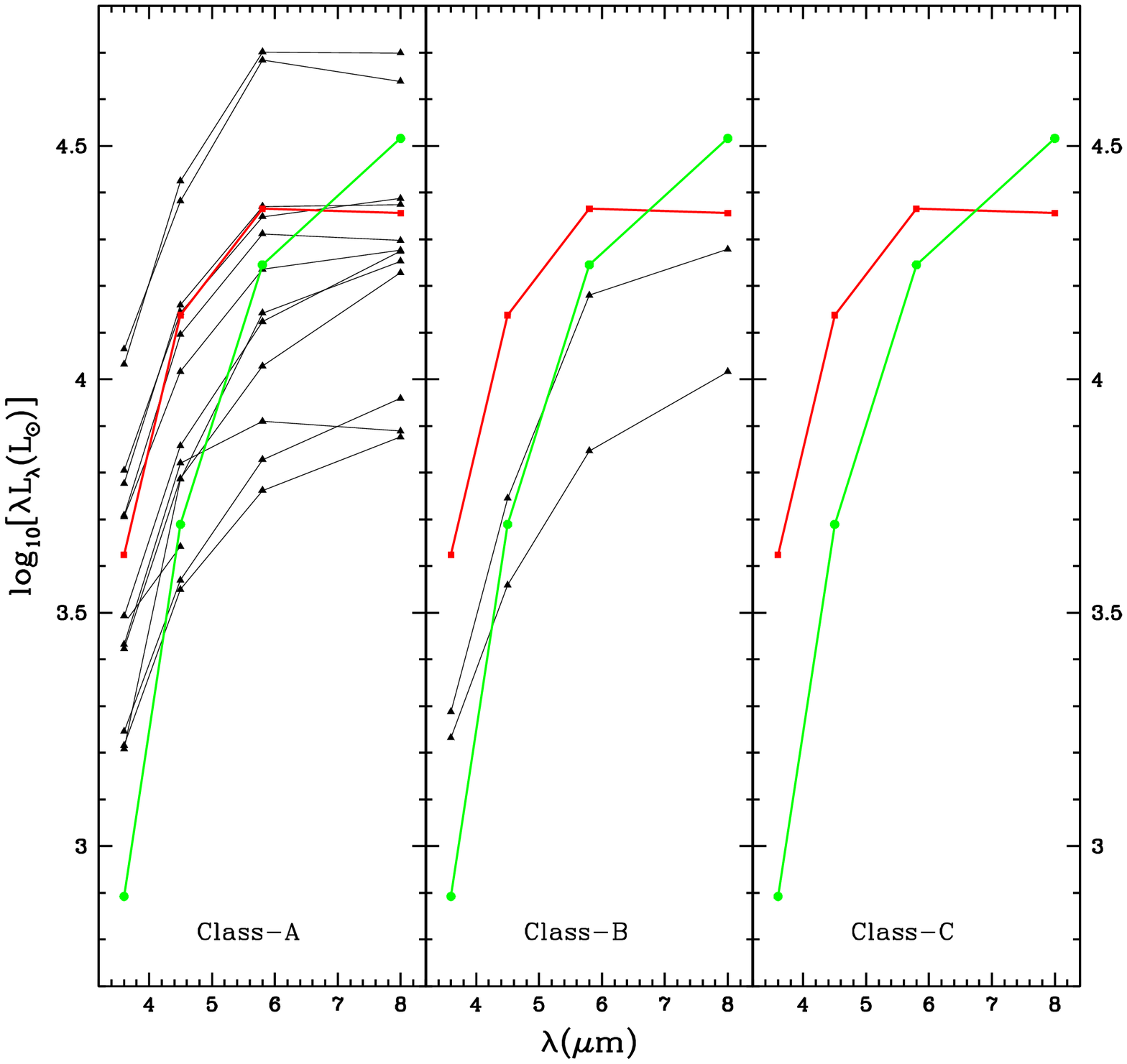}}
\end{center}
\caption{Mid-infrared CMD (\textit{left}) and EAGB SEDs (\textit{right}) for M33. The apparent magnitude at 4.5 $\mu$m is plotted versus $[3.6]-[4.5]$ color for all sources detected in both 3.6 $\mu$m and 4.5 $\mu$m images through PSF-photometry (black dots). For comparison, the positions of the progenitors of NGC~300-2008OT (black circle) and SN~2008S (black square, lower limit in color) are  also shown (\cite{ref:Prieto_2008a,ref:Prieto_2008b}), and the 4.5 $\mu$m absolute magnitude scale is shown on the right. The $[3.6]-[4.5]>1.5$ and $M_{4.5}<-10$ selection for extremely red and bright objects, following the criteria used by~\citet{ref:Thompson_2008}, is shown by the dashed lines. The EAGB candidates that meet these criteria are shown with different symbols sorted according to the stage of the search at which they were identified. The red circles and open red squares indicate Class--A objects identified through PSF and aperture-photometry, the blue triangles indicate Class--B objects, and the green squares indicate Class--C objects (none in this case). Where applicable, the lower limits in color are indicated with arrows. Stars for which only $m_{3.6}$ upper limits could be determined are not shown in the SEDs panel. The SEDs of the SN 2008S (red) and NGC 300-2008OT (green) progenitors are also shown. The 5.8 $\mu$m and 8.0 $\mu$m fluxes were determined through aperture-photometry for the locations identified in the 4.5 $\mu$m image. Due to significant PAH emission in these two bands, we consider the aperture-photometry measurements in these bands less reliable.}
\label{fig:m33_2}
\begin{center}
{\includegraphics[angle=0,width=85mm]{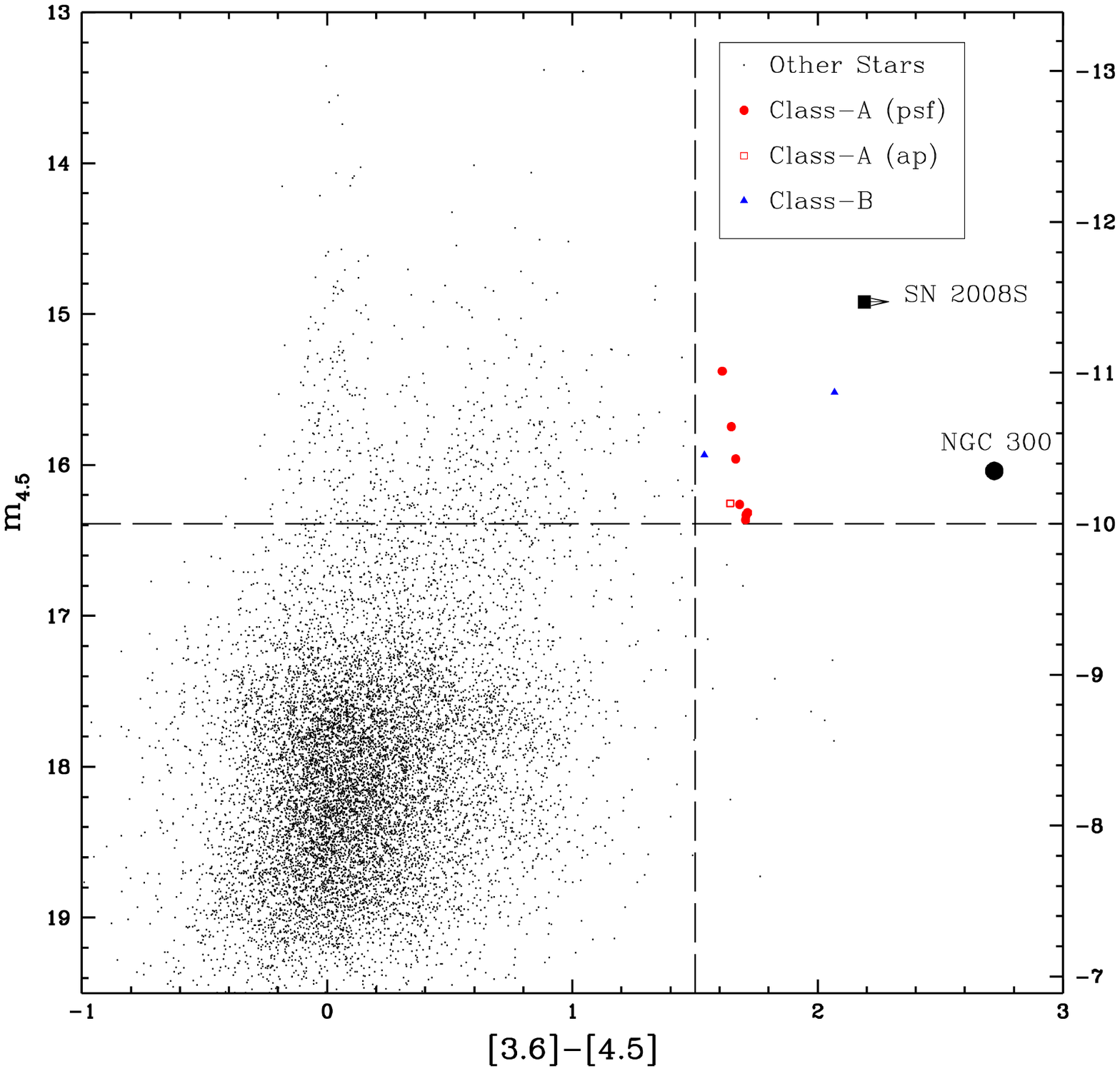}}
{\includegraphics[angle=0,width=85mm]{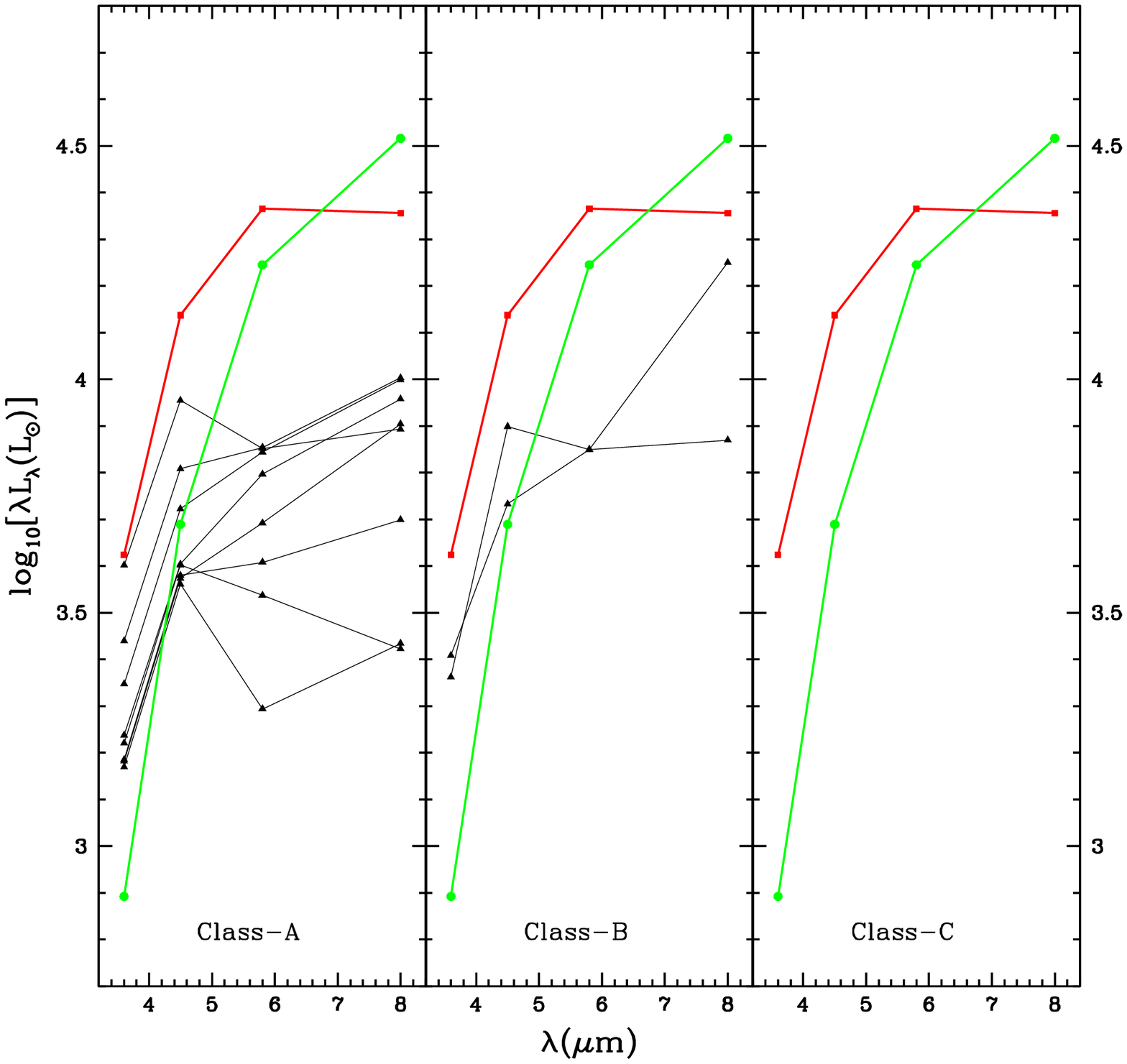}}
\end{center}
\caption{Mid-infrared CMD (\textit{left}) and EAGB SEDs (\textit{right}) for NGC~300. Symbols and colors used here are the same as in Figure~\ref{fig:m33_2}. The SEDs of some fainter sources show a sharp decline at 5.8 $\mu$m before rising again at 8.0 $\mu$m due to PAH dominated background contamination.}
\label{fig:ngc300_2}
\end{figure}

\subsection{NGC~300}

As the host of one of the optical transients, NGC~300-2008OT
\citep[D\,$\simeq1.9$\,Mpc,][]{ref:Gieren_2005} is of particular
interest. Initially, we identify 8 Class--A objects based on
PSF-photometry only (including the NGC~300-2008OT progenitor,
as shown in Figure~\ref{fig:n300_trans}), 1 additional Class--A object
through aperture-photometry, 2 Class--B objects, and 4 Class--C
objects. However, all 4 Class--C objects were rejected as duplicate
detections of Class--A/B objects. Overall, we found 10 candidate
objects in NGC~300 apart from the transient progenitor. The
color determined for each of these 10 objects is above the $2\sigma$
local background limit and none of them is brighter than the
NGC~300-2008OT progenitor and redder than the SN~2008S
progenitor color limit. Table~\ref{table:ngc300_cat} provides a
catalog of the 11,241 sources identified in both bands, and
Table~\ref{table:ngc300_red} lists the candidate EAGB sources.
We estimate the NGC~300-2008OT progenitor properties to be
$M_{3.6}=-7.36$, $M_{4.5}=-10.25$, and $[3.6]-[4.5]=2.89$ 
which agrees with the previous measurement of
$M_{3.6}=-7.63$, $M_{4.5}=-10.39$, and $[3.6]-[4.5]=2.72$
~\citep{ref:Prieto_2008b}. 

\subsection{M81}

As a large, nearby, nearly face-on spiral galaxy at an intermediate
distance (more distant than NGC~300 and closer than NGC~6946), M81
\citep[D\,$\simeq3.6$\,Mpc,][]{ref:Saha_2006} was chosen as a test case
for our improved search method. Initially, we identify 15 Class--A
objects through PSF-photometry only, 9 additional Class--A objects
through aperture-photometry, 15 Class--B objects, and 3 Class--C
objects. However, on visual inspection, we rejected 2 Class--A objects
and 1 Class--C object, leaving 39 candidates, of which only 1
object is brighter than the NGC~300-2008OT progenitor and
redder than the SN~2008S progenitor color limit. For 9 of these
objects we could only determine a $2\sigma$ color limit (4 Class--A, 4
Class--B, 1 Class--C). Table~\ref{table:m81_cat} provides a catalog of
the 6,021 sources identified in both bands, and
Table~\ref{table:m81_red} lists the candidate EAGB sources.

\begin{figure}[p]
\begin{center}
{\includegraphics[angle=0,width=85mm]{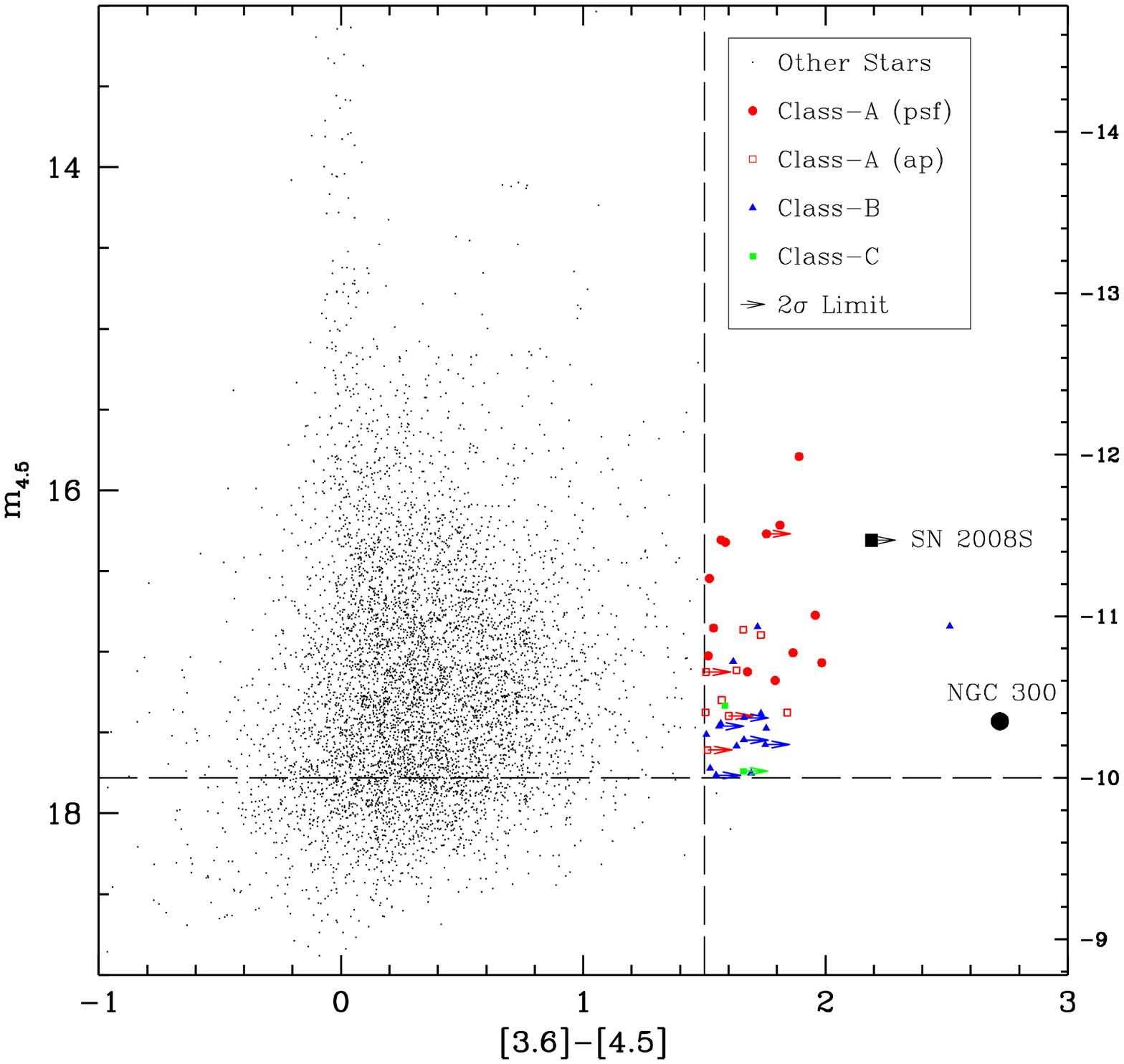}}
{\includegraphics[angle=0,width=85mm]{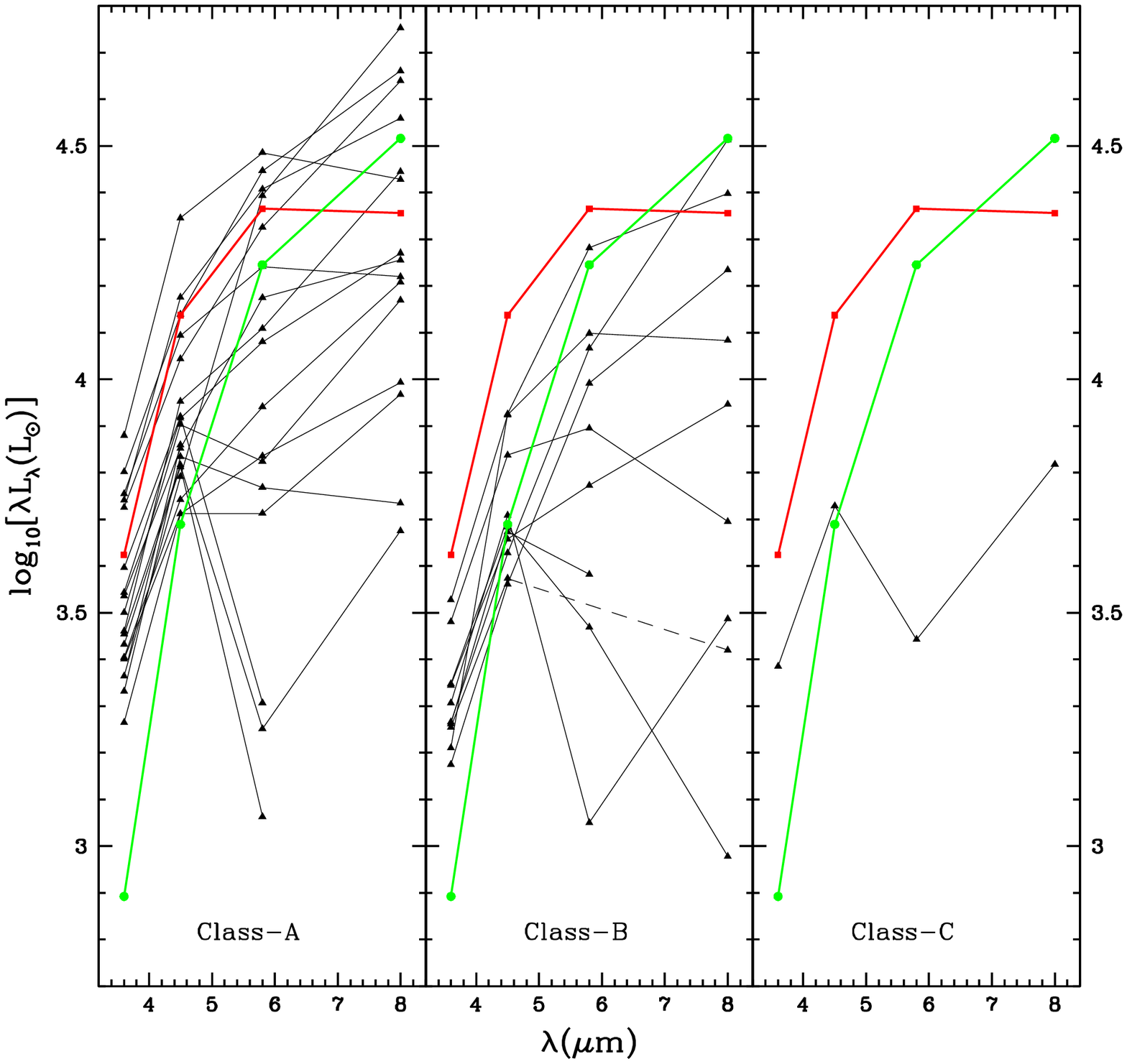}}
\end{center}
\caption{Mid-infrared CMD (\textit{left}) and EAGB SEDs (\textit{right}) for M81. Symbols and colors used here are the same as in Figure~\ref{fig:m33_2}. The SEDs of some fainter sources show a sharp decline at 5.8 $\mu$m before rising again at 8.0 $\mu$m due to PAH dominated background contamination. Sources for which 5.8  $\mu$m and 8.0  $\mu$m measurements could not be obtained at all due to contamination, only the 3.6  $\mu$m and 4.5  $\mu$m measurements are shown on the SEDs. The dashed line indicates an object for which only the 5.8  $\mu$m measurement could not be obtained.}
\label{fig:m81_2}
\begin{center}
{\includegraphics[angle=0,width=85mm]{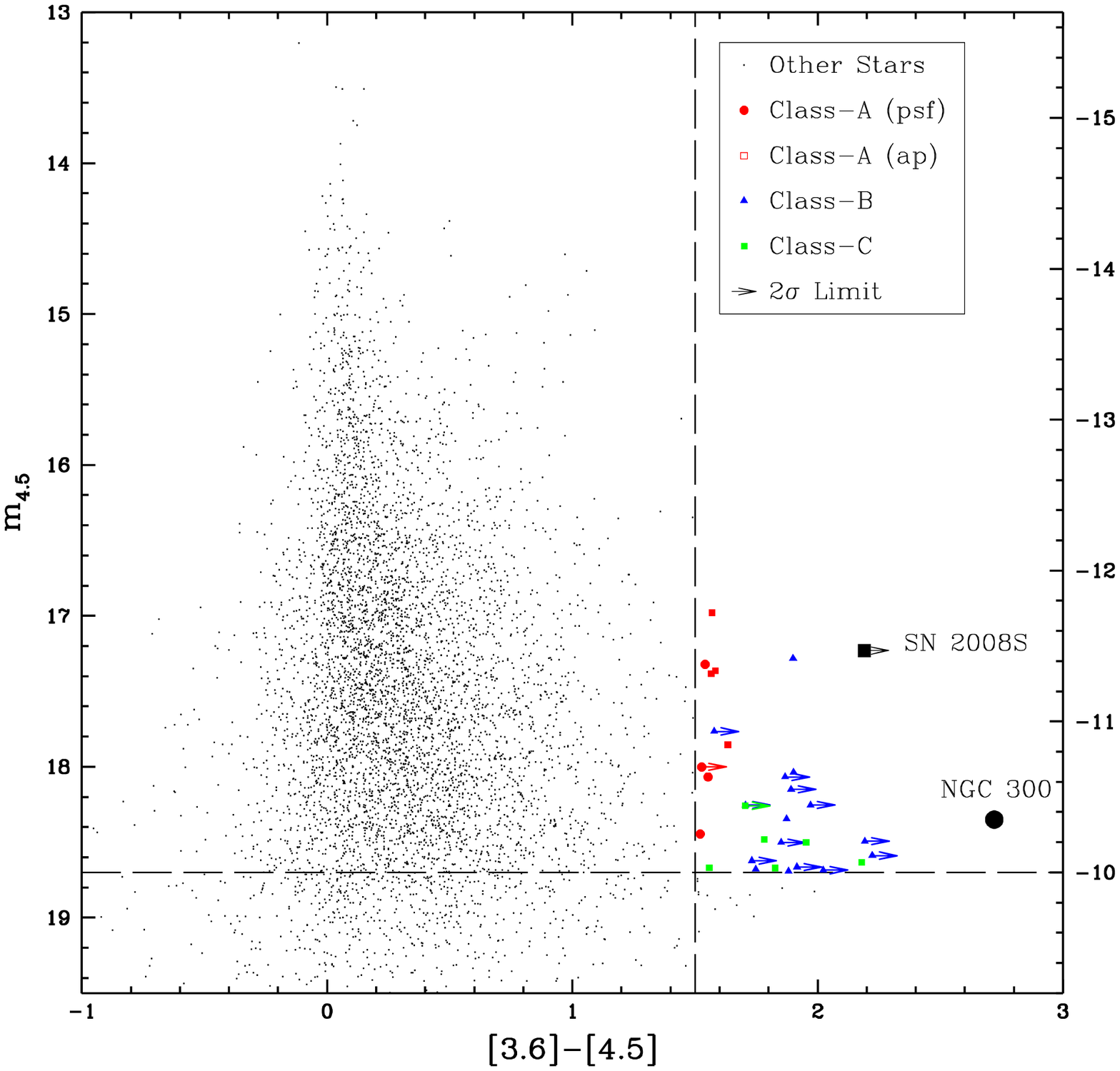}}
{\includegraphics[angle=0,width=85mm]{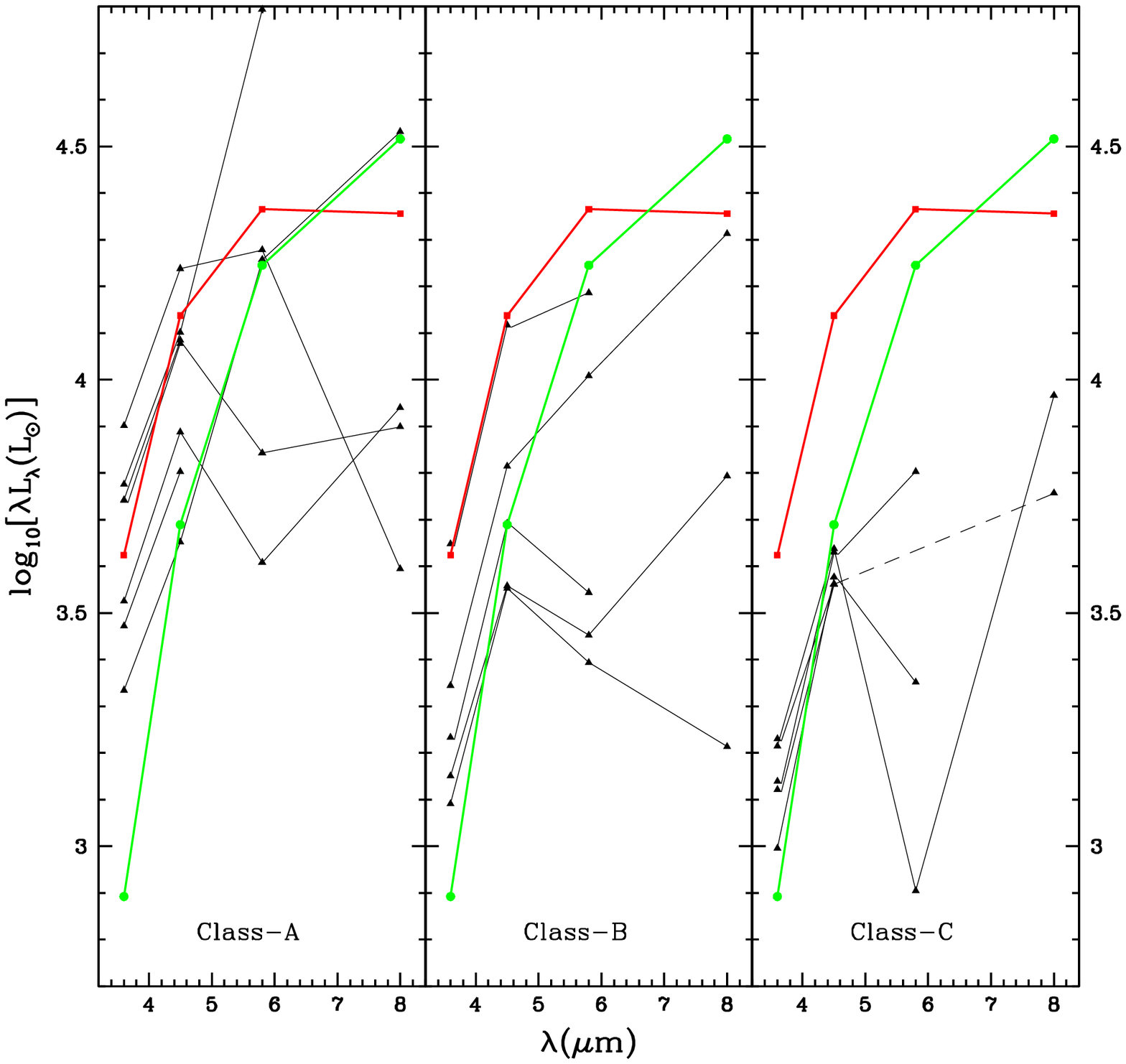}}
\end{center}
\caption{Mid-infrared CMD (\textit{left}) and EAGB SEDs (\textit{right}) for NGC~6946. Symbols and colors used here are the same as in Figure~\ref{fig:m33_2}. The SEDs of some fainter sources show a sharp decline at 5.8 $\mu$m before rising again at 8.0 $\mu$m due to PAH dominated background contamination. Sources for which 5.8  $\mu$m and 8.0  $\mu$m measurements could not be obtained at all due to contamination, only the 3.6  $\mu$m and 4.5  $\mu$m measurements are shown on the SEDs. The dashed line indicates an object for which only the 5.8  $\mu$m measurement could not be obtained.}
\label{fig:ngc6946_2}
\end{figure}

\subsection{NGC~6946}

The host of the SN~2008S event, NGC~6946~\citep[D\,$\simeq5.6$\,Mpc,][]{ref:Sahu_2006} is the most distant of the four galaxies. At this distance, the $M=-10$
absolute magnitude limit is close to the $3\sigma$ detection limit at
4.5 $\mu$m, leading to reduced completeness. The 4.5 $\mu$m  SINGS archival image
of NGC~6946 contains many artifacts that significantly affect the subtracted image. 
For example, bright stars show bright ``halos'' in the 4.5 $\mu$m image that appear 
as rings in the subtracted image. As a
result, most of the candidates identified in the second and
third stages did not pass the visual inspection because they were
clearly associated with artifacts. 

Initially, we identify 5 Class--A objects through PSF-photometry only,
5 additional Class--A objects through aperture-photometry, 38 Class--B
objects (including the SN~2008S progenitor), and 32 Class--C
objects. On visual inspection, we rejected 2 Class--A objects, 21
Class--B objects, and 26 Class--C objects, leaving 30 candidates
apart from the SN~2008S progenitor. None of these new sources are
redder than this progenitor's color limit and brighter than the
NGC~300 transient progenitor. For 13 of these objects we could
only determine a $2\sigma$ color limit (1 Class--A, 11 Class--B, 1
Class--C). Table~\ref{table:ngc6946_cat} provides a catalog of the
5,601 sources identified in both bands, and
Table~\ref{table:ngc6946_red} lists the candidate EAGB sources.

One of the rejected candidates is a slightly shifted duplicate
detection of the SN~2008S event progenitor star. We were unable to
determine if there might be multiple stellar objects blended together
at the image location. This can be further probed in future mid-IR
observations of this galaxy using (warm) \textit{Spitzer}. 
We estimate the SN 2008S progenitor properties to be
$M_{3.6}>-9.09$, $M_{4.5}=-11.28$, and $[3.6]-[4.5]>2.19$ 
which agrees with the previous measurement of
$M_{3.6}>-9.46$, $M_{4.5}=-11.47$, and $[3.6]-[4.5]>2.01$
~\citep{ref:Prieto_2008a}. 

The spatial distribution of AGB and EAGB stars in NGC~6946 is very
different from that of the other three galaxies (see
Figure~\ref{fig:all_sky}). The absence of these stars near the center
of NGC~6946 is real, unlike M81 where we mask the saturated galactic
center. The EAGB candidates also lie on the periphery of the
galaxy near the $R_{25}$ isophotal radius. Brighter and redder EAGBs
are relatively easier to identify through our improved search method
even if they were initially missed due to blending, and it is
highly unlikely that a significant number of EAGB objects in the inner
region of NGC~6946 were missed. This suggests that the
spatial pattern of the star formation history of NGC~6946 is quite
different from the other three galaxies.

%% file: discussion.tex
\section{Discussion}
\label{sec:discussion}

The two supernova-like transient events observed in NGC~300 and
NGC~6946 were fairly luminous, and the mid-IR properties of the
progenitors of both transients indicate the presence of a dusty, warm,
optically thick wind around the
progenitors~\citep{ref:Prieto_2008a,ref:Thompson_2008,ref:Botticella_2009,ref:Prieto_2009,ref:Wesson_2009}. The
extremely red $[3.6]-[4.5]$ colors of the progenitors can only be
explained by total self-obscuration resulting from a period
of circumstellar dust production. Our search for analogs of these progenitors gives us the
opportunity to empirically investigate the nature and origin of such
objects. In this section, we discuss the implications of our
findings. For this discussion, we include the statistics for the LMC
and SMC using the catalogs from the full survey of the LMC
by~\citet{ref:Blum_2006} and the partial survey of the SMC
by~\citet{ref:Bolatto_2007}. Table~\ref{table:mc_red} lists 
the candidate EAGB sources in the LMC and SMC.

First, we attempt to characterize the
detected EAGB population of the target galaxies. Next, we consider how
our results may help improve our understanding of the late-stage evolution
of massive stars. Finally, we motivate a multi-epoch mid-IR survey of
nearby galaxies to produce a complete list of potential progenitors of
future supernova like transient events.

\subsection{The EAGB Population}
\label{sec:rate}

Although the exact number of analogs of the SN~2008S and NGC~300-2008OT
progenitors ($N_{EAGB}$) in these four galaxies is
uncertain due to the absence of absolute quantitative criteria to
identify such objects, it is clear that objects as bright and red as these progenitors are extremely
rare. Following a conservative selection criteria of requiring the
objects to be very red ($[3.6]-[4.5]>1.5$) and bright ($M_{4.5}<-10$)
produces a sample size on the order of tens of candidates per
galaxy. However, following the strictest criteria of requiring the
candidates to be brighter than the NGC~300-2008OT
progenitor and redder than the SN~2008S event progenitor color limit,
we would select no more than 1 candidate object in the LMC, M33 and M81, none in the SMC, and
none other than the 2008 transient progenitors in NGC~300 and NGC~6946. Taking measurement
uncertainties into account can increase the number of objects per galaxy meeting
the strictest criteria by at most 1 more in the LMC, M33, and NGC~300, 0
more in the SMC and M81, and 3 more in NGC~6946.

Simple black-body fits of the EAGB SEDs with an assumed dust emissivity
function going as $\lambda^{-1}$ show that almost all these objects
have a bolometric luminosity of $\sim10^{4}L_{\odot}$ and photospheric
temperatures of $\sim300-600~K$. This indicates that these are very
massive stars embedded in dusty winds, with luminosities and
temperatures very similar to those of the 2008 supernova-like
transient progenitors.

To examine the contamination of the EAGB region of the CMD
by non-EAGB objects, we took a closer look at the LMC objects which
satisfy the $[3.6]-[4.5]>1.5$ and $M_{4.5}<-10$ criteria. The reddest
object at $[3.6]-[4.5]\sim2.5$ is the IRAS source, IRAS
05346-6949~\citep{ref:Elias_1986}, which has been classified as an
``Extremely Red Object'' (ERO) in \citet{ref:Gruendl_2009}. It is 
believed to be an enshrouded supergiant which is not bright in the optical. 
There are IRS spectra for 13 objects in
this (ERO) class, although they are all less luminous
($L\lesssim{10^4}L_{\odot}$) than the 2008 transient progenitors, and 7 of them are
carbon stars~\citep{ref:Gruendl_2008}. If such contamination is
common, then analogs of SN~2008S and NGC~300-2008OT transient
progenitors may be even rarer than our conservative estimates.

To determine the fraction of massive stars in these galaxies that are
analogous to the SN~2008S and NGC~300 transient progenitors, we
first estimate the total number of massive stars in each galaxy by
scaling the \citet{ref:Thompson_2008} estimate of the number of RSG
stars ($N_{RSG}$) in M33. We scale the number using the
\textit{B}-band luminosity estimates from
\citet{ref:Karachentsev_2004}, and find that the fraction is always on
the order of $N_{EAGB}/N_{RSG}\sim10^{-4}$. The constancy of this ratio
means that our efficiency at finding them is nearly constant over a
wide distance range, and that very few of the massive stars in any one
galaxy at any moment have mid-IR colors and brightnesses comparable to
those of the SN~2008S and NGC~300-2008OT progenitors. Table~\ref{table:eagb_fraction} shows the estimated EAGB
counts as a fraction of massive stars in the six galaxies that we
studied. The lower fractions in M81 and NGC~6946 are likely real rather
than a completeness problem. For the SMC we only have a lower bound on
$N_{EAGB}$, $N_{EAGB}/N_{RSG}$, and the ratio relative to M33 
because the ${S^3}MC$ survey of the SMC covered only part of the
galaxy.

\begin{table}[p]
\begin{center}
\caption{EAGB Counts as Fraction of Massive Stars}
\label{table:eagb_fraction}
\begin{tabular}{lrrrrrrrrrrrrr}
\\
\hline 
\hline
\\
\multicolumn{1}{c}{Galaxy} &
\multicolumn{1}{c}{} &
\multicolumn{1}{c}{$N_{EAGB}$} &
\multicolumn{1}{c}{} &
\multicolumn{1}{c}{$M_{B}$} &
\multicolumn{1}{c}{} &
\multicolumn{1}{c}{Luminosity} &
\multicolumn{1}{c}{} &
\multicolumn{1}{c}{$N_{RSG}$} &
\multicolumn{1}{c}{} &
\multicolumn{1}{c}{$N_{EAGB}/N_{RSG}$} &
\multicolumn{1}{c}{} &
\multicolumn{1}{c}{Ratio}
\\
\multicolumn{1}{c}{} &
\multicolumn{1}{c}{} &
\multicolumn{1}{c}{} &
\multicolumn{1}{c}{} &
\multicolumn{1}{c}{} &
\multicolumn{1}{c}{} &
\multicolumn{1}{c}{Ratio to M33} &
\multicolumn{1}{c}{} &
\multicolumn{1}{c}{} &
\multicolumn{1}{c}{} &
\multicolumn{1}{c}{} &
\multicolumn{1}{c}{} &
\multicolumn{1}{c}{to M33}
\\
\hline
\hline
\\
LMC& &9& &$-17.93$& &0.4& &$2.2\times10^4$& &$4.1\times10^{-4}$& &1.11\\
SMC& &$>1$& &$-16.35$& &0.1& &$5.4\times10^3$& &$>1.8\times10^{-4}$& &$>0.49$\\
M33& &20& &$-18.87$& &$\equiv1.0$& &$5.4\times10^4$& &$3.7\times10^{-4}$& &$\equiv1.00$\\
NGC~300& &10& &$-17.92$& &0.4& &$2.2\times10^4$& &$4.5\times10^{-4}$& &1.24\\
M81& &39& &$-21.06$& &7.5& &$4.1\times10^5$& &$1\times10^{-4}$& &0.27\\
NGC~6946& &30& &$-20.86$& &6.3& &$3.4\times10^5$& &$0.9\times10^{-4}$& &0.24\\
\\
\hline
\hline
\\
\end{tabular}
\caption{EAGB Counts as Fraction of AGB Stars}
\label{table:eagb_agb}
\begin{tabular}{lrrrrrrrrrrrrr}
\\
\hline 
\hline
\\
\multicolumn{1}{c}{Galaxy} &
\multicolumn{1}{c}{} &
\multicolumn{1}{c}{$N_{EAGB}$} &
\multicolumn{1}{c}{} &
\multicolumn{1}{c}{$N_{AGB}$ with } &
\multicolumn{1}{c}{} &
\multicolumn{1}{c}{Background} &
\multicolumn{1}{c}{} &
\multicolumn{1}{c}{$N_{AGB}$} &
\multicolumn{1}{c}{} &
\multicolumn{1}{c}{$N_{EAGB}/N_{AGB}$} &
\multicolumn{1}{c}{} &
\multicolumn{1}{c}{Ratio}
\\
\multicolumn{1}{c}{} &
\multicolumn{1}{c}{} &
\multicolumn{1}{c}{} &
\multicolumn{1}{c}{} &
\multicolumn{1}{c}{Contamination} &
\multicolumn{1}{c}{} &
\multicolumn{1}{c}{Contamination} &
\multicolumn{1}{c}{} &
\multicolumn{1}{c}{} &
\multicolumn{1}{c}{} &
\multicolumn{1}{c}{} &
\multicolumn{1}{c}{} &
\multicolumn{1}{c}{to M33}
\\
\hline
\hline
\\
LMC& &9& &400& &$\equiv0$& &400& &0.02& &1.0\\
SMC& &$>1$& &30& &$\equiv0$& &30& &0.03& &1.5\\
M33& &20& &1000& &30& &970& &0.02& &$\equiv1.0$\\
NGC~300& &10& &330& &80& &250& &0.04& &2.0\\
M81& &39& &1310& &510& &800& &0.05& &2.5\\
NGC~6946& &30& &470& &260& &210& &0.14& &7.0\\
\\
\hline
\hline
\\
\end{tabular}
\caption{EAGB Counts and Massive SFR}
\label{table:eagb_sfr}
\begin{tabular}{lrrrrrrrrrrr}
\\
\hline 
\hline
\\
\multicolumn{1}{c}{Galaxy} &
\multicolumn{1}{c}{} &
\multicolumn{1}{c}{$N_{EAGB}$} &
\multicolumn{1}{c}{} &
\multicolumn{1}{c}{$N_{EAGB}$} &
\multicolumn{1}{c}{} &
\multicolumn{1}{c}{log$(L(H{\alpha}))$\tablenotemark{a}} &
\multicolumn{1}{c}{} &
\multicolumn{1}{c}{SFR (H${\alpha}$)\tablenotemark{b}} &
\multicolumn{1}{c}{} &
\multicolumn{1}{c}{SFR (H${\alpha}$)}
\\
\multicolumn{1}{c}{} &
\multicolumn{1}{c}{} &
\multicolumn{1}{c}{} &
\multicolumn{1}{c}{} &
\multicolumn{1}{c}{Ratio to M33} &
\multicolumn{1}{c}{} &
\multicolumn{1}{c}{(erg/s)} &
\multicolumn{1}{c}{} &
\multicolumn{1}{c}{$M_\odot$/yr} &
\multicolumn{1}{c}{} &
\multicolumn{1}{c}{Ratio to M33}
\\
\hline
\hline
\\
LMC& &9& &0.5& &40.5& &0.24& &0.7\\
SMC& &$>1$& &$>0.05$& &39.7& &0.04& &0.1\\
M33& &20& &$\equiv1.0$& &40.6& &0.33& &$\equiv1.0$\\
NGC~300& &10& &0.5& &40.2& &0.13& &0.4\\
M81& &39& &2.0& &40.8& &0.46& &1.4\\
NGC~6946& &30& &1.5& &41.4& &2.05& &6.3\\
\\
\hline
\hline
\\
\end{tabular}
\end{center}
\tablenotetext{a}{Estimates from \citet{ref:Kennicutt_2008} scaled for distances used in this paper.}
\tablenotetext{b}{Estimated using Equation 2 of \citet{ref:Lee_2009}.}
\end{table}

\begin{figure}[p]
\begin{center}
{\includegraphics[angle=0,width=180mm]{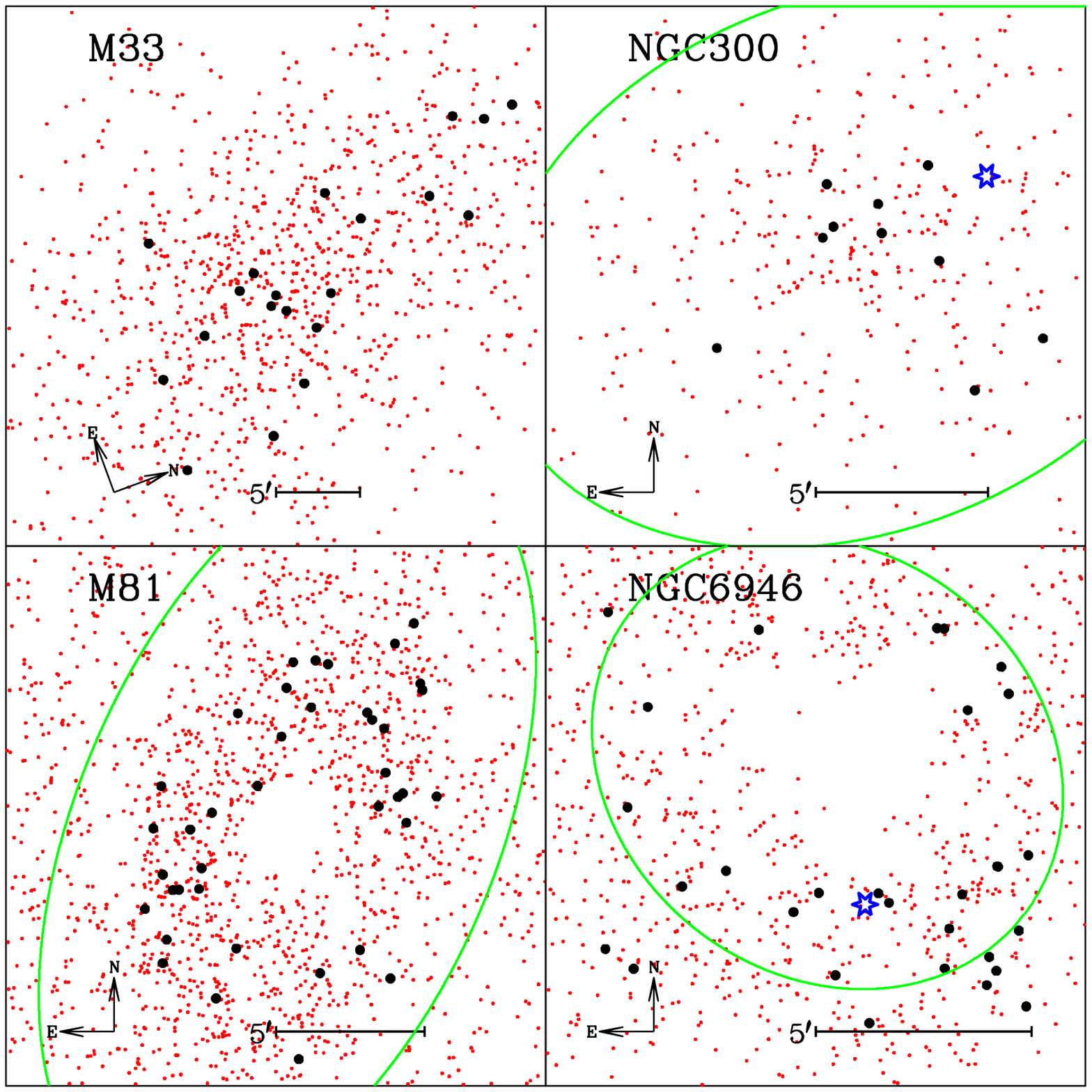}}
\end{center}
\caption{Distribution of AGB stars (red dots) and EAGB stars (black filled circles), and the two 2008 transient locations (starred blue symbols) in the galaxies. The image scales and directions are indicated in each panel, as well as the $R_{25}$ ellipse (green). For M33, the $R_{25}$ ellipse lies outside of our angular selection region. The empty region at the center of M81 is due to the image mask that we used for the brightest, nearly saturated, central region of the galaxy. However, the notable absence of any AGB and EAGB candidates towards the central region of NGC~6946 is not artificial, as discussed in Section~\ref{sec:population}. We estimate that for NGC~300, M81, and NGC~6946, 90\%, 70\%, and 50\% of our angular selection region is inside the $R_{25}$ ellipse (minus the M81 mask and the ``empty'' central region of NGC~6946). We use these sky area estimates when scaling the extragalactic contamination from SDWFS.}
\label{fig:all_sky}
\end{figure}

\citet{ref:Thompson_2008} proposed that the very red and bright
candidates are the most massive AGB stars in a late stage of
evolution and thus constitute a natural extension of the AGB
branch. Under this assumption, the ratio of the number of EAGB and
massive AGB stars in a galaxy at a given time should depend on the
star formation history of the galaxy, since the lifetime of
$\sim2-8M_{\odot}$ stars spans $\sim100$Myr--$1$Gyr. Based on the
structure of the M33 CMD and the detection limits for the more
distant galaxies, we used the criteria of
\begin{equation} 
-9.3 < M_{4.5} + 3([3.6]-[4.5]) < -7.3 
\end{equation}
and
\begin{equation}
-10.2 < M_{4.5} - 0.55([3.6]-[4.5]) < -11.6
\end{equation}
to select the bulk of the AGB sequence while minimizing the effects of incompleteness for the more distant galaxies. We also required that the objects lie within the $R_{25}$ isophotal radius of the galaxies (see Figure~\ref{fig:all_sky}).

\begin{figure}[p]
\begin{center}
{\includegraphics[angle=0,width=180mm]{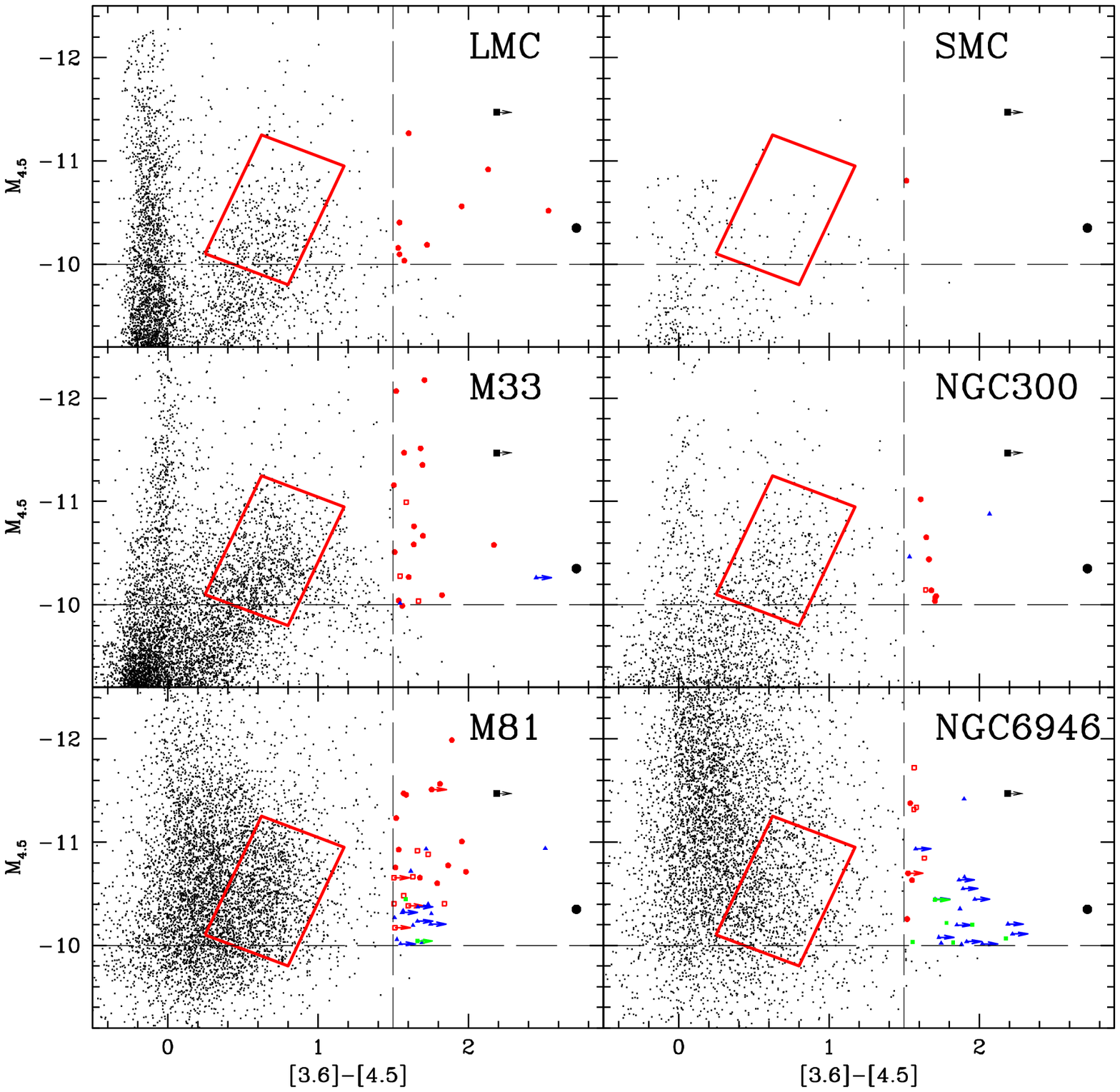}}
\end{center}
\caption{Mid-infrared CMDs for the six galaxies. Symbols and colors used here are same as in Figure~\ref{fig:m33_2}, and the AGB region is shown in red. The small number of bright objects in the SMC CMD is largely due to the ${S^3}MC$ survey~\citep{ref:Bolatto_2007} covering only portions of the SMC. The AGB regions of the M81 and NGC~6946 CMDs contain significant extragalactic contamination.}
\label{fig:all_cmd}
\end{figure}

Figure~\ref{fig:all_cmd} shows the CMDs zoomed in
to highlight the AGB and EAGB stars, and Figure~\ref{fig:all_sky}
shows the sky distribution of the luminous red AGB and EAGB candidate
objects. The small number of bright objects in the SMC CMD is largely
due to the ${S^3}MC$ survey~\citep{ref:Bolatto_2007} covering only portions
of the SMC. While we can clearly see the AGB sequence in M33 and
NGC~300, contamination by extragalactic sources is a problem for M81
and NGC~6946. We correct for this contamination using the 10 square
degree Spitzer Deep Wide Field
Survey~\citep[SDWFS,][]{ref:Ashby_2009}. While there is no significant
extragalactic contamination to the EAGB sources,
Table~\ref{table:eagb_agb} shows that the contamination correction to
the AGB sample is important. We ignore the possibility of foreground
contamination of the AGB region given the rarity of luminous
AGB stars in our galaxy and our small fields of view.

The ratios of the numbers of EAGB to luminous red AGB stars, in the
LMC, SMC, and the four galaxies that we studied, are
$N_{EAGB}/N_{AGB}=$ $0.01$, $0.015$, $0.02$, $0.04$, $0.05$, and
$0.14$, respectively. There is worrisome trend in the ratios with increasing
distance. It may be partly due to decreasing completeness of the AGB
population relative to that for the EAGB population with increasing
distance. The AGB candidates are selected from the catalog of
mid-IR sources that are detected in both 3.6 $\mu$m and 4.5 $\mu$m,
and for galaxies at greater distances we are losing AGB
stars at the bottom of our selection box and potentially
overestimating the extragalactic contamination. This can lead us to
over-subtract the contamination and underestimate $N_{AGB}$. On the
other hand, the EAGB candidates are detected using our improved
methods, and thus include objects detected only at 4.5 $\mu$m or in
the $[3.6]-[4.5]$ differenced image as well as those detected in both
IRAC bands. This allows us to detect EAGB candidates that would
otherwise be missed due to crowding and thus makes our EAGB count
relatively more complete at greater distances when compared to the
completeness of the AGB star count.

Table~\ref{table:eagb_agb} summarizes the EAGB counts as fraction of
AGB stars. If all these galaxies had the same relative star formation
history, the $N_{EAGB}/N_{AGB}$ ratio should be identical for all
galaxies given fixed completeness. Alternatively, the various steps we
used for finding the EAGB stars are improving our completeness for EAGB stars
at the distance of NGC~6946 by a factor of $\sim7$. This may well be
the case, since if we only used the DAOPHOT catalog of this galaxy, we
would find only 4 EAGB candidates instead of 30. Thus, for
NGC~6946 and to a lesser extent M81, our completeness for AGB stars is
limited by the shallow depth of the SINGS observations.

Next we investigate whether the numbers of EAGB candidates correlate
with the current SFR of these galaxies. We use
the $L(H{\alpha})$ estimates from \citet{ref:Kennicutt_2008} and
estimate the SFR using the results from \citet{ref:Lee_2009}. There may be
a weak correlation between $N_{EAGB}$ and SFR (H$\alpha$), although
the correlation clearly does not hold for NGC~6946. The discrepancy in
this case is severe, and corresponds to finding only $\sim20\%$ of the
EAGB stars if viewed as a completeness problem. It seems more likely
that the correlation is coincidental and that the EAGB stars must generally 
be less massive than the $\gtrsim15$ $M_{\odot}$ stars responsible for the 
$H{\alpha}$ emissions.

The EAGB candidate objects are expected to be
very faint in the optical even if an optical counterpart exists. As such, we
need very deep optical catalogs to search for optical counterparts. Of the four
galaxies that we studied, only M33 has a publicly available deep optical
catalog. \citet{ref:Thompson_2008} reported that no optical counterparts 
of the M33 EAGB candidate sources was found.

\subsection{Understanding the Evolution of Massive Stars}
\label{sec:evolution}

As discussed in \citet{ref:Thompson_2008}, there are three classes of
explanations for the two transient events detected in 2008: explosive white 
dwarf formation, low-luminosity supernova (normal Fe core or electron capture SN), or a massive star
transient. \citet{ref:Prieto_2008a} and \citet{ref:Thompson_2008}
argue for the electron capture SN interpretation, and this is
supported by \citet{ref:Botticella_2009}. \citet{ref:Pumo_2009} argue
for this scenario using a parametric approach, demonstrating that such
supernovae can be explained in terms of electron capture SN from
super-AGB progenitors. \citet{ref:Prieto_2009} also found that the
mid-IR spectra of the outburst resembled that of a proto-planetary
nebula rather than a massive star outburst. \citet{ref:Kashi_2010} suggest 
that a mass transfer episode from an extreme asymptotic giant branch star 
to a main sequence companion powered NGC~3002008-OT. \citet{ref:Smith_2009} and
\citet{ref:Bond_2009} argue for a massive star outburst because of the
optical spectra of the transients, but \citet{ref:Prieto_2008a} note that
such optical spectral characteristics are also found in
proto-planetary nebula. \citet{ref:Smith_2009b} suggest that the 
two transient events are possibly linked to LBV eruptions,
and are similar to SN~2009ip and the 2009 optical transient in 
UGC~2773. However, both of these events discovered in 2009 have optically 
luminous progenitors, unlike the progenitors of SN~2008S and NGC~300-2008OT.
\citet{ref:Bonanos_2009} suggested that objects
similar to the two 2008 transient progenitors are related to supergiant Be stars.
 However, whether a link between LBVs and sgBe stars exists is not known at this point.

While the nature of the two 2008 transient progenitors remains
debated, our search for analogs of such objects has now empirically
shown that such extremely red and luminous objects are truly rare. 
We estimate this number to be on the order
of few tens per large galaxy, but it can be as low as $\sim1$ if
selected by the strictest criteria of requiring the objects to be
brighter than the NGC~300 transient progenitor and redder than the
SN~2008S progenitor. \citet{ref:Thompson_2008} estimated
the duration of the dust obscured phase of EAGB stars to be on the
order of $\sim10^{4}$ years, a time scale very similar to that of the
onset of carbon burning in relatively low mass massive stars.

Late stage evolution of stars in this mass range, around
$\sim6-8M_{\odot}$, is very difficult to model
\citep[e.g.,][]{ref:Miyaji_1980,ref:Siess_2007,ref:Poelarends_2008}. \citet{ref:Pumo_2009}
argue that for very rapid mass loss rates, a completely self-obscuring
shell can even be produced on the order of $\sim150$ years prior to an
electron capture SN explosion. Such an ultra-short duration for the
dust obscured phase would make it surprising to find any true analogs
of the SN~2008S or NGC~3002008OT progenitors in our small sample
of galaxies.

We cannot definitively claim that any of our EAGB candidates
are truly analogous to the two 2008 transient 
progenitors. However, we have empirically shown that even using
relatively shallow archival data for relatively distant galaxies, we
can successfully identify the dusty self-obscured stellar
population. Using a combination of conventional PSF and
aperture-photometry methods along with innovative use of difference
imaging methods, we can overcome crowding issues for galaxies at
greater distances as well. If another SN~2008S like transient is
to occur in these galaxies, it is likely to be one of the sources we
identified here. While the ordering of our search from very conventional 
(Class--A) to unconventional (Class--C) appears to suggest that the 
wavelength-differenced approach is of limited use, it is in fact very 
valuable for confirming candidates and becomes increasingly effective 
at greater distances. It is important when using it to note the manner in 
which the resulting changes in image statistics somewhat confuse standard source 
identification codes. 

This study is also a significant expansion of the sample of galaxies
with detailed mid-IR stellar catalogs. Previously, full catalogs
existed only for the LMC~\citep{ref:Blum_2006} and
M33~\citep{ref:McQuinn_2007,ref:Thompson_2008}, with partial catalogs
for the SMC~\citep{ref:Bolatto_2007} and
M31~\citep{ref:Mould_2008}. Our new catalogs can be used to study
various stellar populations in nearby galaxies over a wide range of
distances. We have also demonstrated that an innovative application of
difference imaging can be used to remove crowding confusion for
distant galaxies when searching for objects with significant color
excesses. The improved data analysis method used in this study can be
applied to study other stellar populations as well.

\subsection{Motivations for Future Observations}
\label{sec:complete}

Significant uncertainties regarding the nature of these transient
events and their progenitors remain. A simple step is to
determine whether the SN~2008S or NGC~300-2008OT progenitors survived their
transients. The transient light curves have now likely faded, based on 
the \citet{ref:Botticella_2009} and \citet{ref:Bond_2009} light-curves, to 
the point where the continued existence of a progenitor can be observed, 
although this will likely require \textit{Spitzer}
observations to constrain a return to a self-obscured phase if optical
and near-IR searches fail. Another important step is to properly
survey the AGB/EAGB populations of all nearby ($D<10$~Mpc)
galaxies. Such a survey would have two goals. First, it would detect
and characterize the progenitors of any future, similar events, and
generally constrain the fraction of supernovae and massive star
transients associated with self-obscured phases of stellar
evolution more systematically. \citet{ref:Kochanek_2008} and 
\citet{ref:Smartt_2009} point out that there is a deficit 
of massive SN progenitors. One solution would be to obscure the 
progenitors and such \textit{Spitzer} searches will greatly help to constrain 
this possibility, albeit only for self-obscured sources as compared 
to stars behind dense foreground dust. 
Second, it would characterize these phases of stellar
evolution more generally, particularly since treatments of the AGB
phase are an increasingly important problem in models of galaxy
evolution~\citep[e.g.,][]{ref:Marcillac_2006,ref:Santini_2009,ref:Conroy_2010}. Given catalogs
of the AGB stars in galaxies with a broad range of physical
properties, the scaling of AGB stars and their mid-IR emission due to
dust absorption could be empirically calibrated as a function of
galaxy star formation rates and metallicity.

For the more distant galaxies, M81 and NGC~6946, we relied on the SINGS
observations, which were not designed to characterize individual
stars. As a result, they are shallower than desired for this
purpose. For example, the observational depth makes it difficult to
count the AGB stars in these galaxies reliably and to properly control
for contamination by extragalactic sources. We emphasize that source
confusion for fully self-obscured stars is not a primary limitation and can
be overcome even in the more distant galaxies using our new analysis
methods based on difference imaging between the 3.6 $\mu$m and 4.5
$\mu$m wavelengths.

A particularly attractive approach to obtaining such deeper catalogs
is to carry out the observations as part of a sparse variability
survey. Little is known about mid-IR variability of such rare,
massive stars, and the variability provides another means of
identifying AGB stars in the presence of confusion, again through 
the use of difference imaging. In~\citet{ref:Thompson_2008}, for example, 
we found that variability increased along
the AGB sequence except for the small population of EAGB stars, but
the statistics were so limited that it was hard to draw general
conclusions.

%% file: conclusion.tex
\section{Conclusions}
\label{sec:conclusion}

We carried out a systematic mid-IR photometric search for massive,
luminous, self-obscured stars in four nearby galaxies combined with 
existing data for the LMC and SMC. We use a
combination of conventional PSF and aperture-photometry techniques
along with an innovative application of image subtraction. We
investigate the population of SN~2008S-like transient event progenitor
analogs in these 6 galaxies. We report
catalogs of mid-IR sources in three new galaxies (NGC~300, M81, and
NGC~6946) and candidate extreme AGB stars in all 6.

Using our methods, bright and red extreme AGB stars can be inventoried
to $D\lesssim10$ Mpc despite \textit{Spitzer}'s relatively poor angular
resolution. The biggest current problem is that the archival data is
insufficiently deep: even with our innovative ``band-subtraction''
technique, we simply need more photons to detect these extremely red
stars in the more distant galaxies.  A future multi-epoch survey using
(warm) \textit{Spitzer} could identify all EAGB candidates in
nearby galaxies, characterize their variability, and pin down the
contribution of these partly obscured stars to galaxy SEDs as a function of wavelength.

Finally, we again emphasize the point made by~\citet{ref:Thompson_2008}. 
Stars analogous to the progenitors of the SN~2008S and the NGC~300 transients 
are truly rare in all galaxies.  At any moment there appears to be only $\sim1$ 
true analog, and up to $\sim10$ given a more liberal selection criterion, per galaxy. 
While completeness problems due to the limited depth of the archival data make
 it impossible to give exact scalings, they represent roughly $2 \times 10^{-4}$ 
of the red super giant population, $\sim 10^{-2}$ of the AGB population, and 
appear at a rate of order $50$ EAGB stars per unit star formation 
($M_\odot$~year$^{-1}$) using the liberal criteria (and an order of magnitude 
fewer if we use the more conservative one). Clarifying these scalings with stellar 
mass, SFR and metallicity requires larger and deeper surveys of nearby galaxies
than can be accomplished with warm \textit{Spitzer} or eventually with JWST.

\acknowledgments

We thank the referee for helpful comments,
Szymon Kozlowski for helping us estimate extragalactic
contamination using the SDWFS data, and Janice Lee for helpful
discussions. We extend our gratitude to the SINGS Legacy Survey and LVL Survey for
making their data publicly available. This research has made use of
NED, which is operated by the JPL and Caltech, under contract with
NASA and the HEASARC Online Service, provided by NASA's GSFC. RK and
KZS are supported in part by NSF grant AST-0707982. JLP acknowledges support from NASA through Hubble
Fellowship grant HF-51261.01-A awarded by the STScI, which
is operated by AURA, Inc. for NASA, under contract NAS 5-26555.  KZS, CSK and TAT are supported in
part by NSF grant AST-0908816. TAT is supported in part by an Alfred
P. Sloan Foundation Fellowship. JFB is supported by NSF CAREER grant
PHY-0547102.

%% file: appendix.tex
\begin{appendix}
\label{appendix}

\section{LMC and SMC}
\label{app:mc}

\begin{center}
\begin{longtable}{lcccrrrrrrrrrrrr}
\caption{Photometry for the EAGBs in LMC\tablenotemark{c} and SMC\tablenotemark{d}}
\label{table:mc_red}
\\
\hline 
\hline
\\
\multicolumn{1}{c}{Galaxy} &
\multicolumn{1}{c}{RA} &
\multicolumn{1}{c}{Dec} &
\multicolumn{1}{c}{} &
\multicolumn{1}{c}{$[3.6]$} &
\multicolumn{1}{c}{} &
\multicolumn{1}{c}{$[4.5]$} &
\multicolumn{1}{c}{} &
\multicolumn{1}{c}{Color} &
\multicolumn{1}{c}{} &
\multicolumn{1}{c}{$[5.8]$} &
\multicolumn{1}{c}{} &
\multicolumn{1}{c}{$[8.0]$} \\
\multicolumn{1}{c}{} &
\multicolumn{1}{c}{(deg)} &
\multicolumn{1}{c}{(deg)} &
\multicolumn{1}{c}{} &
\multicolumn{1}{c}{(mag)} &
\multicolumn{1}{c}{} &
\multicolumn{1}{c}{(mag)} &
\multicolumn{1}{c}{} &
\multicolumn{1}{c}{} &
\multicolumn{1}{c}{} &
\multicolumn{1}{c}{(mag)} &
\multicolumn{1}{c}{} &
\multicolumn{1}{c}{(mag)} \\
\\
\hline
\hline
\\
\endfirsthead
\multicolumn{3}{c}
{{\tablename\ \thetable{} -- continued from previous page}}
\\
\hline 
\hline
\\
\multicolumn{1}{c}{Galaxy} &
\multicolumn{1}{c}{RA} &
\multicolumn{1}{c}{Dec} &
\multicolumn{1}{c}{} &
\multicolumn{1}{c}{$[3.6]$} &
\multicolumn{1}{c}{} &
\multicolumn{1}{c}{$[4.5]$} &
\multicolumn{1}{c}{} &
\multicolumn{1}{c}{Color} &
\multicolumn{1}{c}{} &
\multicolumn{1}{c}{$[5.8]$} &
\multicolumn{1}{c}{} &
\multicolumn{1}{c}{$[8.0]$} \\
\multicolumn{1}{c}{} &
\multicolumn{1}{c}{(deg)} &
\multicolumn{1}{c}{(deg)} &
\multicolumn{1}{c}{} &
\multicolumn{1}{c}{(mag)} &
\multicolumn{1}{c}{} &
\multicolumn{1}{c}{(mag)} &
\multicolumn{1}{c}{} &
\multicolumn{1}{c}{} &
\multicolumn{1}{c}{} &
\multicolumn{1}{c}{(mag)} &
\multicolumn{1}{c}{} &
\multicolumn{1}{c}{(mag)} \\
\\
\hline
\hline
\\
\endhead
\\
\hline
\hline
\multicolumn{4}{r}{{Continued on next page}}
\\ 
\hline
\hline
\\
\endfoot
\endlastfoot
LMC &81.694135 &$-$68.813077& &8.74& &7.13& &1.61& &5.79& &4.37\\
LMC &83.559766 &$-$69.788996& &10.42& &7.88& &2.21& &5.67& &3.65\\
LMC &78.003216 &$-$70.540071& &9.94& &8.22& &1.72& &6.82& &5.44\\
LMC &79.801060 &$-$69.152029& &9.80& &7.84& &1.96& &6.17& &5.01\\
LMC &80.552349 &$-$67.975638& &9.54& &8.00& &1.54& &6.56& &5.20\\
LMC &82.726151 &$-$68.574502& &9.78& &8.24& &1.54& &7.10& &5.98\\
LMC &83.332279 &$-$69.695785& &9.85& &8.31& &1.54& &7.14& &6.05\\
LMC &84.932592 &$-$69.642732& &9.62& &7.48& &2.14& &5.80& &4.63\\
LMC &87.485476 &$-$70.886629& &9.94& &8.37& &1.57& &7.08& &5.81\\
SMC &16.246012 &$-$72.147362& &9.61& &8.09& &1.52& & \dots & &7.88\\
\\
\hline
\hline
\\
\tablenotetext{c}{From the catalog published in~\citet{ref:Blum_2006}.}
\tablenotetext{d}{From the catalog published in~\citet{ref:Bolatto_2007}.}
\end{longtable}
\end{center}

\section{M33}
\label{app:m33}

\begin{center}
\begin{longtable}{lcccrrrrrrrrrrrr}
\caption{Photometry for the 20 EAGBs in M33}
\label{table:m33_red}
\\
\hline 
\hline
\\
\multicolumn{1}{c}{RA} &
\multicolumn{1}{c}{} &
\multicolumn{1}{c}{Dec} &
\multicolumn{1}{c}{} &
\multicolumn{1}{c}{$[3.6]$} &
\multicolumn{1}{c}{} &
\multicolumn{1}{c}{$[4.5]$} &
\multicolumn{1}{c}{} &
\multicolumn{1}{c}{Color} &
\multicolumn{1}{c}{} &
\multicolumn{1}{c}{$[5.8]$} &
\multicolumn{1}{c}{} &
\multicolumn{1}{c}{$[8.0]$} \\
\multicolumn{1}{c}{(deg)} &
\multicolumn{1}{c}{} &
\multicolumn{1}{c}{(deg)} &
\multicolumn{1}{c}{} &
\multicolumn{1}{c}{(mag)} &
\multicolumn{1}{c}{} &
\multicolumn{1}{c}{(mag)} &
\multicolumn{1}{c}{} &
\multicolumn{1}{c}{} &
\multicolumn{1}{c}{} &
\multicolumn{1}{c}{(mag)} &
\multicolumn{1}{c}{} &
\multicolumn{1}{c}{(mag)} \\
\\
\hline
\hline
\\
\endfirsthead
\multicolumn{3}{c}
{{\tablename\ \thetable{} -- continued from previous page}}
\\
\hline 
\hline
\\
\multicolumn{1}{c}{RA} &
\multicolumn{1}{c}{} &
\multicolumn{1}{c}{Dec} &
\multicolumn{1}{c}{} &
\multicolumn{1}{c}{$[3.6]$} &
\multicolumn{1}{c}{} &
\multicolumn{1}{c}{$[4.5]$} &
\multicolumn{1}{c}{} &
\multicolumn{1}{c}{Color} &
\multicolumn{1}{c}{} &
\multicolumn{1}{c}{$[5.8]$} &
\multicolumn{1}{c}{} &
\multicolumn{1}{c}{$[8.0]$} \\
\multicolumn{1}{c}{(deg)} &
\multicolumn{1}{c}{} &
\multicolumn{1}{c}{(deg)} &
\multicolumn{1}{c}{} &
\multicolumn{1}{c}{(mag)} &
\multicolumn{1}{c}{} &
\multicolumn{1}{c}{(mag)} &
\multicolumn{1}{c}{} &
\multicolumn{1}{c}{} &
\multicolumn{1}{c}{} &
\multicolumn{1}{c}{(mag)} &
\multicolumn{1}{c}{} &
\multicolumn{1}{c}{(mag)} \\
\\
\hline
\hline
\\
\endhead
\\
\hline
\hline
\multicolumn{4}{r}{{Continued on next page}}
\\ 
\hline
\hline
\\
\endfoot
\endlastfoot
23.55626& &30.55206& &14.43& &12.72& &1.71& &11.27& &10.29\\ 
23.55223& &30.90565& &15.07& &13.39& &1.68& &12.10& &11.11\\ 
23.34230& &30.64595& &15.00& &13.43& &1.57& &12.15& &11.07\\ 
23.45476& &30.85700& &16.49& &14.32& &2.17& &12.67& &11.41\\ 
23.40428& &30.51731& &15.24& &13.55& &1.69& &12.25& &11.03\\ 
23.29855& &30.59900& &16.47& &14.91& &1.56& &13.62& &12.35\\ 
23.56806& &30.87753& &16.63& &14.81& &1.82& &13.77& &12.80\\ 
23.55592& &30.93663& &16.40& &14.86& &1.54& &13.45& &12.14\\ 
23.39692& &30.67733& &15.25& &13.74& &1.51& &12.44& &11.35\\ 
23.53805& &30.73258& &15.78& &14.14& &1.64& &12.72& &11.36\\ 
23.44840& &30.65119& &14.35& &12.83& &1.52& &11.32& &10.45\\ 
23.43436& &30.57101& &15.93& &14.23& &1.70& &13.25& &12.32\\ 
23.46795& &30.61916& &15.95& &14.32& &1.63& &12.95& &11.47\\ 
23.49594& &30.75659& &16.24& &14.63& &1.61& &13.82& &12.90\\ 
23.43886& &30.64303& &15.90& &14.39& &1.51& &13.41& &12.49\\ 
23.29759& &30.50737& &15.50& &13.91& &1.59& &12.55& &11.27\\ 
23.42809& &30.70271& &16.17& &14.62& &1.55& &13.06& &11.26\\ 
23.42741& &30.65543& &16.54& &14.87& &1.67& & \dots & & \dots \\ 
23.48104& &30.63851& &$>$17.09& &14.64& &$>$2.45& &12.92& &11.83\\ 
23.49179& &30.82798& &16.43& &14.89& &1.54& &13.41& &12.00\\ 
\\
\hline
\hline
\\
\end{longtable}
\end{center}

\newpage
\section{NGC~300}
\label{app:ngc300}

\begin{table}[h]
\begin{center}
\caption{MIR Catalog for 11,241 Point Sources in NGC~300}
\label{table:ngc300_cat}
\begin{tabular}{lrrrrrrrrrrr}
\\
\hline 
\hline
\\
\multicolumn{1}{c}{RA} &
\multicolumn{1}{c}{} &
\multicolumn{1}{c}{Dec} &
\multicolumn{1}{c}{} &
\multicolumn{1}{c}{$[3.6]$} &
\multicolumn{1}{c}{$\sigma_{3.6}$} &
\multicolumn{1}{c}{} &
\multicolumn{1}{c}{$[4.5]$} &
\multicolumn{1}{c}{$\sigma_{4.5}$} &
\multicolumn{1}{c}{} &
\multicolumn{1}{c}{Color}
\\
\multicolumn{1}{c}{(deg)} &
\multicolumn{1}{c}{} &
\multicolumn{1}{c}{(deg)} &
\multicolumn{1}{c}{} &
\multicolumn{1}{c}{(mag)} &
\multicolumn{1}{c}{} &
\multicolumn{1}{c}{} &
\multicolumn{1}{c}{(mag)} &
\multicolumn{1}{c}{} &
\multicolumn{1}{c}{} &
\multicolumn{1}{c}{} \\
\\
\hline
\hline
\\
13.85991& &$-$37.57455& &11.50 &0.02& &11.50 &0.03& &0.01\\
13.62310& &$-$37.81105& &12.04 &0.02& &12.03 &0.02& &0.01\\
13.87644& &$-$37.56594& &12.04 &0.02& &12.05 &0.01& &$-$0.00\\
13.71941& &$-$37.81866& &12.73 &0.01& &12.68 &0.02& &0.04\\
13.83688& &$-$37.80313& &12.93 &0.02& &12.88 &0.03& &0.05\\
\dots    & & \dots    & & \dots & \dots &  & \dots & \dots & & \dots \\
\\
\hline
\hline
\\
\end{tabular}
\end{center}
\end{table}

\begin{center}
\begin{longtable}{lcccrrrrrrrrrrrr}
\caption{Photometry for the 10 EAGBs in NGC~300}
\label{table:ngc300_red}
\\
\hline 
\hline
\\
\multicolumn{1}{c}{RA} &
\multicolumn{1}{c}{} &
\multicolumn{1}{c}{Dec} &
\multicolumn{1}{c}{} &
\multicolumn{1}{c}{$[3.6]$} &
\multicolumn{1}{c}{} &
\multicolumn{1}{c}{$[4.5]$} &
\multicolumn{1}{c}{} &
\multicolumn{1}{c}{Color} &
\multicolumn{1}{c}{} &
\multicolumn{1}{c}{$[5.8]$} &
\multicolumn{1}{c}{} &
\multicolumn{1}{c}{$[8.0]$} \\
\multicolumn{1}{c}{(deg)} &
\multicolumn{1}{c}{} &
\multicolumn{1}{c}{(deg)} &
\multicolumn{1}{c}{} &
\multicolumn{1}{c}{(mag)} &
\multicolumn{1}{c}{} &
\multicolumn{1}{c}{(mag)} &
\multicolumn{1}{c}{} &
\multicolumn{1}{c}{} &
\multicolumn{1}{c}{} &
\multicolumn{1}{c}{(mag)} &
\multicolumn{1}{c}{} &
\multicolumn{1}{c}{(mag)} \\
\\
\hline
\hline
\\
\endfirsthead
\multicolumn{3}{c}
{{\tablename\ \thetable{} -- continued from previous page}}
\\
\hline 
\hline
\\
\multicolumn{1}{c}{RA} &
\multicolumn{1}{c}{} &
\multicolumn{1}{c}{Dec} &
\multicolumn{1}{c}{} &
\multicolumn{1}{c}{$[3.6]$} &
\multicolumn{1}{c}{} &
\multicolumn{1}{c}{$[4.5]$} &
\multicolumn{1}{c}{} &
\multicolumn{1}{c}{Color} &
\multicolumn{1}{c}{} &
\multicolumn{1}{c}{$[5.8]$} &
\multicolumn{1}{c}{} &
\multicolumn{1}{c}{$[8.0]$} \\
\multicolumn{1}{c}{(deg)} &
\multicolumn{1}{c}{} &
\multicolumn{1}{c}{(deg)} &
\multicolumn{1}{c}{} &
\multicolumn{1}{c}{(mag)} &
\multicolumn{1}{c}{} &
\multicolumn{1}{c}{(mag)} &
\multicolumn{1}{c}{} &
\multicolumn{1}{c}{} &
\multicolumn{1}{c}{} &
\multicolumn{1}{c}{(mag)} &
\multicolumn{1}{c}{} &
\multicolumn{1}{c}{(mag)} \\
\\
\hline
\hline
\\
\endhead
\\
\hline
\hline
\multicolumn{4}{r}{{Continued on next page}}
\\ 
\hline
\hline
\\
\endfoot
\endlastfoot
13.60942& &$-$37.72041& &18.03& &16.32& &1.71& &15.49& &14.28\\
13.80838& &$-$37.72509& &18.04& &16.33& &1.71& &15.28& &13.76\\
13.74378& &$-$37.67188& &18.07& &16.36& &1.71& &16.27& &14.94\\
13.65096& &$-$37.74543& &17.94& &16.26& &1.68& &15.66& &14.97\\
13.71002& &$-$37.65558& &17.63& &15.96& &1.67& &14.90& &13.53\\
13.73724& &$-$37.66650& &17.39& &15.75& &1.64& &14.87& &13.52\\
13.74127& &$-$37.64604& &16.99& &15.38& &1.61& &14.88& &13.79\\
13.67270& &$-$37.68299& &17.90& &16.26& &1.64& &15.02& &13.63\\
13.70787& &$-$37.66964& &17.59& &15.52& &2.07& &14.88& &12.90\\
13.67960& &$-$37.63695& &17.47& &15.93& &1.54& &14.88& &13.85\\
\\
\hline
\hline
\\
\end{longtable}
\end{center}

\newpage
\section{M81}
\label{app:m81}
\begin{table}[h]
\begin{center}
\caption{MIR Catalog for 6,021 Point Sources in M81}
\label{table:m81_cat}
\begin{tabular}{lrrrrrrrrrrr}
\\
\hline 
\hline
\\
\multicolumn{1}{c}{RA} &
\multicolumn{1}{c}{} &
\multicolumn{1}{c}{Dec} &
\multicolumn{1}{c}{} &
\multicolumn{1}{c}{$[3.6]$} &
\multicolumn{1}{c}{$\sigma_{3.6}$} &
\multicolumn{1}{c}{} &
\multicolumn{1}{c}{$[4.5]$} &
\multicolumn{1}{c}{$\sigma_{4.5}$} &
\multicolumn{1}{c}{} &
\multicolumn{1}{c}{Color}
\\
\multicolumn{1}{c}{(deg)} &
\multicolumn{1}{c}{} &
\multicolumn{1}{c}{(deg)} &
\multicolumn{1}{c}{} &
\multicolumn{1}{c}{(mag)} &
\multicolumn{1}{c}{} &
\multicolumn{1}{c}{} &
\multicolumn{1}{c}{(mag)} &
\multicolumn{1}{c}{} &
\multicolumn{1}{c}{} &
\multicolumn{1}{c}{} \\
\\
\hline
\hline
\\
149.19633& &69.12408& &11.26 &0.05& &11.30 &0.04& &$-$0.04\\
149.32484& &69.02681& &12.74 &0.05& &12.80 &0.04& &$-$0.06\\
149.30643& &69.05417& &12.86 &0.03& &12.85 &0.03& &0.02\\
148.72163& &69.06341& &12.93 &0.04& &12.88 &0.04& &0.05\\
149.21478& &69.12849& &13.96 &0.02& &12.96 &0.04& &1.00\\
\dots    & & \dots    & & \dots & \dots &  & \dots & \dots & & \dots \\
\\
\hline
\hline
\\
\end{tabular}
\end{center}
\end{table}
\begin{center}
\begin{longtable}{lcccrrrrrrrrrrrr}
\caption{Photometry for the 39 EAGBs in M81}
\label{table:m81_red}
\\
\hline 
\hline
\\
\multicolumn{1}{c}{RA} &
\multicolumn{1}{c}{} &
\multicolumn{1}{c}{Dec} &
\multicolumn{1}{c}{} &
\multicolumn{1}{c}{$[3.6]$} &
\multicolumn{1}{c}{} &
\multicolumn{1}{c}{$[4.5]$} &
\multicolumn{1}{c}{} &
\multicolumn{1}{c}{Color} &
\multicolumn{1}{c}{} &
\multicolumn{1}{c}{$[5.8]$} &
\multicolumn{1}{c}{} &
\multicolumn{1}{c}{$[8.0]$} \\
\multicolumn{1}{c}{(deg)} &
\multicolumn{1}{c}{} &
\multicolumn{1}{c}{(deg)} &
\multicolumn{1}{c}{} &
\multicolumn{1}{c}{(mag)} &
\multicolumn{1}{c}{} &
\multicolumn{1}{c}{(mag)} &
\multicolumn{1}{c}{} &
\multicolumn{1}{c}{} &
\multicolumn{1}{c}{} &
\multicolumn{1}{c}{(mag)} &
\multicolumn{1}{c}{} &
\multicolumn{1}{c}{(mag)} \\
\\
\hline
\hline
\\
\endfirsthead
\multicolumn{3}{c}
{{\tablename\ \thetable{} -- continued from previous page}}
\\
\hline 
\hline
\\
\multicolumn{1}{c}{RA} &
\multicolumn{1}{c}{} &
\multicolumn{1}{c}{Dec} &
\multicolumn{1}{c}{} &
\multicolumn{1}{c}{$[3.6]$} &
\multicolumn{1}{c}{} &
\multicolumn{1}{c}{$[4.5]$} &
\multicolumn{1}{c}{} &
\multicolumn{1}{c}{Color} &
\multicolumn{1}{c}{} &
\multicolumn{1}{c}{$[5.8]$} &
\multicolumn{1}{c}{} &
\multicolumn{1}{c}{$[8.0]$} \\
\multicolumn{1}{c}{(deg)} &
\multicolumn{1}{c}{} &
\multicolumn{1}{c}{(deg)} &
\multicolumn{1}{c}{} &
\multicolumn{1}{c}{(mag)} &
\multicolumn{1}{c}{} &
\multicolumn{1}{c}{(mag)} &
\multicolumn{1}{c}{} &
\multicolumn{1}{c}{} &
\multicolumn{1}{c}{} &
\multicolumn{1}{c}{(mag)} &
\multicolumn{1}{c}{} &
\multicolumn{1}{c}{(mag)} \\
\\
\hline
\hline
\\
\endhead
\\
\hline
\hline
\multicolumn{4}{r}{{Continued on next page}}
\\ 
\hline
\hline
\\
\endfoot
\endlastfoot
148.67606 & &69.14224 & &18.03 & &16.22 & &1.81 & &14.88 & &13.51\\
148.65425 & &69.08266 & &18.97 & &17.18 & &1.79 & &14.91 & &13.03\\
148.89754 & &69.11637 & &17.68 & &15.79 & &1.89 & &14.68 & &13.84\\
148.72737 & &68.98086 & &18.54 & &17.02 & &1.52 & &15.46 & &14.27\\
148.83730 & &68.98399 & &19.05 & &17.07 & &1.98 & &16.47 & &15.58\\
149.06765 & &69.03034 & &17.88 & &16.31 & &1.57 & &14.78 & &13.26\\
149.00651 & &69.07363 & &18.87 & &17.01 & &1.86 & &14.22 & &12.23\\
148.74550 & &69.07730 & &17.91 & &16.32 & &1.59 & &14.98 & &13.63\\
149.02259 & &69.04263 & &18.39 & &16.85 & &1.54 & &17.63 & & \dots \\
148.88968 & &69.14354 & &18.73 & &16.77 & &1.96 & &15.62 & &13.80\\
148.93513 & &69.08881 & &18.80 & &17.12 & &1.68 & &18.24 & & \dots \\
149.09832 & &69.06480 & &18.07 & &16.55 & &1.52 & &15.08 & &13.31\\
149.02652 & &69.03100 & &$>$18.03 & &16.27 & &$>$1.76 & &15.54 & &14.57\\
148.84383 & &69.15902 & &18.52 & &16.86 & &1.66 & &15.69 & &14.24\\
149.07671 & &69.00262 & &18.63 & &16.90 & &1.73 & &16.33 & &14.49\\
148.71896 & &69.16830 & &19.22 & &17.38 & &1.84 & &16.31 & &14.92\\
148.75517 & &69.12566 & &18.74 & &17.11 & &1.63 & &17.77 & &15.72\\
148.76269 & &69.12973 & &18.87 & &17.30 & &1.57 & &16.04 & &14.39\\
148.96612 & &69.12932 & &18.88 & &17.37 & &1.51 & &16.61 & &14.99\\
148.99972 & &68.96981 & &$>$19.00 & &17.40 & &$>$1.60 & &16.89 & & \dots \\
148.96812 & &68.99771 & &$>$19.12 & &17.61 & &$>$1.51 & &16.09 & &14.55\\
148.73406 & &69.09608 & &$>$18.63 & &17.12 & &$>$1.51 & &16.34 & &15.44\\
149.04063 & &69.06424 & &$>$19.31 & &17.76 & &$>$1.55 & &18.52 & &16.15\\
148.77491 & &68.99690 & &19.44 & &17.75 & &1.69 & &15.92 & &14.33\\
148.87053 & &68.93584 & &19.25 & &17.72 & &1.53 & & \dots & &16.36\\
148.87911 & &69.15796 & &19.22 & &17.59 & &1.63 & &15.73 & &13.63\\
149.08604 & &69.08849 & &$>$19.21 & &17.55 & &$>$1.66 & &16.00 & &14.52\\
148.70745 & &69.08449 & &19.02 & &17.51 & &1.51 & &16.46 & &15.05\\
148.82431 & &69.15687 & &19.23 & &17.47 & &1.76 & &16.94 & & \dots \\
148.73618 & &69.12076 & &$>$19.02 & &17.46 & &$>$1.56 & & \dots & & \dots \\
148.68902 & &69.17968 & &19.01 & &17.44 & &1.57 & &17.22 & &17.46\\
149.11110 & &69.01974 & &$>$19.07 & &17.41 & &$>$1.66 & &16.73 & &15.42\\
149.05787 & &69.03054 & &19.11 & &17.38 & &1.73 & &18.27 & &16.20\\
149.08270 & &68.98944 & &18.68 & &17.06 & &1.62 & &16.16 & &15.67\\
148.85093 & &69.13267 & &18.56 & &16.84 & &1.72 & &15.65 & &14.70\\
148.67919 & &69.14593 & &19.36 & &16.84 & &2.52 & &15.19 & &13.92\\
148.70187 & &69.06804 & &$>$19.32 & &17.57 & &$>$1.75 & &16.04 & &14.21\\
148.71479 & &69.08243 & &$>$19.40 & &17.74 & &$>$1.66 & &17.31 & &16.86\\
149.08331 & &69.03896 & &18.92 & &17.33 & &1.59 & &17.29 & &15.36\\
\\
\hline
\hline
\\
\end{longtable}
\end{center}
\newpage
\section{NGC~6946}
\label{app:ngc6946}

\begin{table}[h]
\begin{center}
\caption{MIR Catalog for 5,601 Point Sources in NGC~6946}
\label{table:ngc6946_cat}
\begin{tabular}{lrrrrrrrrrrr}
\\
\hline 
\hline
\\
\multicolumn{1}{c}{RA} &
\multicolumn{1}{c}{} &
\multicolumn{1}{c}{Dec} &
\multicolumn{1}{c}{} &
\multicolumn{1}{c}{$[3.6]$} &
\multicolumn{1}{c}{$\sigma_{3.6}$} &
\multicolumn{1}{c}{} &
\multicolumn{1}{c}{$[4.5]$} &
\multicolumn{1}{c}{$\sigma_{4.5}$} &
\multicolumn{1}{c}{} &
\multicolumn{1}{c}{Color}
\\
\multicolumn{1}{c}{(deg)} &
\multicolumn{1}{c}{} &
\multicolumn{1}{c}{(deg)} &
\multicolumn{1}{c}{} &
\multicolumn{1}{c}{(mag)} &
\multicolumn{1}{c}{} &
\multicolumn{1}{c}{} &
\multicolumn{1}{c}{(mag)} &
\multicolumn{1}{c}{} &
\multicolumn{1}{c}{} &
\multicolumn{1}{c}{} \\
\\
\hline
\hline
\\
308.77542& &60.06477& &12.94 &0.03& &12.78 &0.03& &0.16\\
308.72086& &60.17000& &13.09 &0.03& &13.20 &0.03& &$-$0.12\\
308.80408& &60.04553& &13.53 &0.03& &13.50 &0.03& &0.04\\
308.69391& &60.11113& &13.57 &0.03& &13.51 &0.03& &0.06\\
308.77582& &60.09848& &13.66 &0.03& &13.51 &0.03& &0.15\\
\dots    & & \dots    & & \dots & \dots &  & \dots & \dots & & \dots \\
\\
\hline
\hline
\\
\end{tabular}
\end{center}
\end{table}

\begin{center}
\begin{longtable}{lcccrrrrrrrrrrrr}
\caption{Photometry for the 30 EAGBs in NGC~6946}
\label{table:ngc6946_red}
\\
\hline 
\hline
\\
\multicolumn{1}{c}{RA} &
\multicolumn{1}{c}{} &
\multicolumn{1}{c}{Dec} &
\multicolumn{1}{c}{} &
\multicolumn{1}{c}{$[3.6]$} &
\multicolumn{1}{c}{} &
\multicolumn{1}{c}{$[4.5]$} &
\multicolumn{1}{c}{} &
\multicolumn{1}{c}{Color} &
\multicolumn{1}{c}{} &
\multicolumn{1}{c}{$[5.8]$} &
\multicolumn{1}{c}{} &
\multicolumn{1}{c}{$[8.0]$} \\
\multicolumn{1}{c}{(deg)} &
\multicolumn{1}{c}{} &
\multicolumn{1}{c}{(deg)} &
\multicolumn{1}{c}{} &
\multicolumn{1}{c}{(mag)} &
\multicolumn{1}{c}{} &
\multicolumn{1}{c}{(mag)} &
\multicolumn{1}{c}{} &
\multicolumn{1}{c}{} &
\multicolumn{1}{c}{} &
\multicolumn{1}{c}{(mag)} &
\multicolumn{1}{c}{} &
\multicolumn{1}{c}{(mag)} \\
\\
\hline
\hline
\\
\endfirsthead
\multicolumn{3}{c}
{{\tablename\ \thetable{} -- continued from previous page}}
\\
\hline 
\hline
\\
\multicolumn{1}{c}{RA} &
\multicolumn{1}{c}{} &
\multicolumn{1}{c}{Dec} &
\multicolumn{1}{c}{} &
\multicolumn{1}{c}{$[3.6]$} &
\multicolumn{1}{c}{} &
\multicolumn{1}{c}{$[4.5]$} &
\multicolumn{1}{c}{} &
\multicolumn{1}{c}{Color} &
\multicolumn{1}{c}{} &
\multicolumn{1}{c}{$[5.8]$} &
\multicolumn{1}{c}{} &
\multicolumn{1}{c}{$[8.0]$} \\
\multicolumn{1}{c}{(deg)} &
\multicolumn{1}{c}{} &
\multicolumn{1}{c}{(deg)} &
\multicolumn{1}{c}{} &
\multicolumn{1}{c}{(mag)} &
\multicolumn{1}{c}{} &
\multicolumn{1}{c}{(mag)} &
\multicolumn{1}{c}{} &
\multicolumn{1}{c}{} &
\multicolumn{1}{c}{} &
\multicolumn{1}{c}{(mag)} &
\multicolumn{1}{c}{} &
\multicolumn{1}{c}{(mag)} \\
\\
\hline
\hline
\\
\endhead
\\
\hline
\hline
\multicolumn{4}{r}{{Continued on next page}}
\\ 
\hline
\hline
\\
\endfoot
\endlastfoot
308.74430& &60.09641& &$>$19.53& &18.00& &$>$1.53& &17.22& &16.32\\
308.68564& &60.05348& &19.62& &18.07& &1.55& & \dots & & \dots \\
308.57731& &60.18055& &18.86& &17.32& &1.54& &14.83& &12.83\\
308.85752& &60.17550& &19.97& &18.45& &1.52& &16.17& &14.50\\
308.58661& &60.11402& &19.49& &17.86& &1.63& &17.80& &15.98\\
308.58305& &60.19101& &18.94& &17.36& &1.58& &17.21& &16.08\\
308.60922& &60.17430& &18.55& &16.98& &1.57& &16.12& &16.84\\
308.79663& &60.11229& &18.95& &17.38& &1.57& & \dots & & \dots \\
308.62378& &60.08993& &$>$19.96& &18.25& &$>$1.71& &17.22& &15.98\\
308.88858& &60.21206& &$>$20.71& &18.69& &$>$2.02& & \dots & &15.49\\
308.58582& &60.11384& &$>$20.04& &18.15& &$>$1.89& &17.51& &16.51\\
308.62727& &60.07459& &$>$20.35& &18.50& &$>$1.85& &17.27& &16.87\\
308.87304& &60.13669& &$>$20.22& &18.25& &$>$1.97& &17.75& &20.10\\
308.67866& &60.10364& &$>$19.94& &18.07& &$>$1.87& &17.36& &14.64\\
308.83064& &60.10620& &$>$20.68& &18.49& &$>$2.19& & \dots & & \dots \\
308.62745& &60.20580& &$>$20.58& &18.66& &$>$1.92& & \dots & & \dots \\
308.77135& &60.20534& &19.94& &18.04& &1.90& &16.79& &15.05\\
308.72470& &60.10376& &19.18& &17.28& &1.90& &16.35& & \dots \\
308.56433& &60.05990& &20.57& &18.69& &1.88& &18.33& &17.80\\
308.63279& &60.20597& &20.22& &18.35& &1.87& &17.96& & \dots \\
308.71184& &60.07200& &20.43& &18.68& &1.75& &18.18& &16.35\\
308.67038& &60.09997& &$>$19.35& &17.77& &$>$1.58& &16.71& &16.09\\
308.56994& &60.08913& &$>$20.81& &18.59& &$>$2.22& &17.32& &17.35\\
308.59289& &60.07903& &$>$20.35& &18.62& &$>$1.73& &16.87& &18.58\\
308.61373& &60.10318& &$>$19.96& &18.26& &$>$1.70& & \dots & &18.88\\
308.56247& &60.11828& &20.81& &18.63& &2.18& &18.44& & \dots \\
308.88990& &60.08202& &20.45& &18.50& &1.95& &17.31& &19.94\\
308.86798& &60.07440& &20.50& &18.67& &1.83& & \dots & &16.44\\
308.59478& &60.06813& &20.26& &18.48& &1.78& &19.55& &15.92\\
308.58739& &60.07366& &20.23& &18.67& &1.56& &20.11& & \dots \\
\\
\hline
\hline
\\
\end{longtable}
\end{center}

\end{appendix}